\documentclass[a4paper,11pt]{article}
\pdfoutput=1 

\usepackage{jcappub} 

\usepackage[T1]{fontenc} 
\usepackage{multirow}
\usepackage{multicol}

\begin{document}

\title{Observational constraints on 
$f(Q,T)$ gravity from the mass–radius relation and stability of compact stars}                     

\author[a]{S. K. Maurya,}
\author[a]{Abdul Aziz,}
\author[b]{Ksh. Newton Singh,}
\author[c]{G. Mustafa,}
\author[d]{Y. Sekhmani,}
\author[e]{Saibal Ray}

\affiliation[a]{Department of Mathematics and Physical Sciences, University of Nizwa, Nizwa 616, Oman}
\affiliation[b]{Department of Physics and Astrophysics, University of Delhi, New Delhi-110007, India.}
\affiliation[c]{Astrophysics Research Centre, School of Mathematics, Statistics and Computer Science, University of KwaZulu-Natal, Private Bag X54001, Durban 4000, South Africa}
\affiliation[d]{Center for Theoretical Physics, Khazar University, 41 Mehseti Street, Baku, AZ1096, Azerbaijan.}
\affiliation[e]{Centre for Cosmology, Astrophysics and Space Science (CCASS), GLA University, Mathura 281406, Uttar Pradesh, India}

\emailAdd{sunil@unizwa.edu.om}
\emailAdd{azizmail2012@gmail.com}
\emailAdd{ntnphy@gmail.com}
\emailAdd{gmustafa3828@gmail.com} 
\emailAdd{sekhmaniyassine@gmail.com}
\emailAdd{saibal.ray@gla.ac.in}

\date{\today}
\abstract{In this investigation we examine the astrophysical consequences of the influence of  pressure anisotropy on the physical properties of observed pulsars within the background of $f(Q,T)$ gravity by choosing a specific form  $f(Q, T)=\psi_1\, Q + \psi_2 T$, where $\psi_1$ and $\psi_2$ are the model parameters. Initially,  we solve the modified field equations for anisotropic stellar configurations by assuming the physically valid metric potential along with  anisotropic function for the distribution of the interior matter.  We test the derived gravitational model subject to various stability conditions to confirm physically existence of compact stars within the $f(Q,T)$ gravity context. We analyze thoroughly the influence of anisotropy on the effective density, pressure and mass-radius relation of the stars. The present inspection of the model implies  that the current gravitational models are non-singular and able to justify for the occurrence of observed pulsars with masses exceeding 2 $M_{\odot}$ as well as masses fall in the {\em mass gap} regime,  in particular merger events like  GW190814. The predicted radii for the observed stars of different masses fall within the  range \{10.5 km, 14.5 km\} for $\psi_1\leq 1.05$ whereas the radius of PSR J074+6620 is predicted to fall within \{13.09 km, 14.66 km\} which is in agreement with the predicted radii range \{11.79 km, 15.01 km\} as can be found in the recent literature.}


\maketitle
\flushbottom

\section{Introduction}\label{sec:1}

~~~~~~~~~~By the late twentieth century, astronomers first perceived that cosmic expansion was not merely continuing but also accelerating. This has become such a discovery that completely surprised  the theorists in the field of astrophysics and cosmology. Subsequently, the common assumption that gravity’s persistent attraction would decelerate and ultimately reverse the Universe’s growth became questionable. So, it would be anticipated that the mutual pull of all matter, governed by gravity, must eventually stops the expansion and turns it  into a contraction. However, experimental evidences in the field of astronomy and cosmology such as the luminosity distances of Type Ia supernovae \cite{Planck:2015fie,DES:2016jjg}, CMB (cosmic background background) \cite{Turk:2010ah} and surveys of large scale structure \cite{SDSS:2005xqv,WMAP:2012nax} have confirmed from all fronts that the Universe is expanding with an increasing rate.  Hence, from theses results it can be concurred that the predictions of classical gravitational theory contradicts the experimental evidences. As a consequence, therefore, we must acknowledge the fact that the present physical laws governing the gravity are incomplete and need to pursue innovative theoretical constructions which would be consistent with the observationally obtained verified results.

Various gravitational theories have been attempted to describe the gravity in terms of different geometrical quantities in order to match the theoretical results with the observational datasets. For instance, gravity is represented by a manifestation of spacetime curvature in the framework of General Relativity (GR) and alternative gravity theories such as $f(R)$ gravity \cite{BeltranJimenez:2018vdo,BeltranJimenez:2017tkd,BeltranJimenez:2019tme} where, in general, the Ricci scalar $R$ in the action is substituted with some functional forms of it.  In addition to curvature there are another two basic geometrical entities, such as torsion and non-metricity, by which gravity theories equivalent to GR can be formulated mathematically as suggested by various contemporary researchers.  With this regard, the Levi-Civita connection as well as the metric tensor play an important role in describing the gravity in background of a pseudo Riemannian spacetime. In the metric teleparallel formalism \cite{Maluf2013,Bahamonde2021} of gravity, the \textit{Weitzenböck connection} and vanishing curvature with imposition of extra constraint, $\nabla_\mu g_{\alpha \beta}$ implies that the gravity can be governed by the torsion tensor. On the other hand, the gravity in a torsion-free region can be governed by the non-metricity tensor. In particular, this is known as the symmetric teleparallel gravity \cite{Nester1999,Adak2006}. When this type of gravity have similar form of field equations as in GR, we call it symmetric teleparallel equivalent of general relativity (STEGR) \cite{Jimenez2018,Jimenez2020}. 

By contracting indices of non-metricity tensor we obtain the non-metricity scalar ($Q$) which takes a role in the action of STEGR equivalent to the role of the Ricci scalar in the case of an action in GR. Further extension to the STEGR is made as this faces cosmology constant problem in connection to the $\Lambda$CDM model. This extension to the STEGR is comparable to $f(R)$ gravity and known as $f(Q)$ gravity \cite{BeltranJimenez:2017tkd} where $f(Q)$ is a functional form of $Q$. Without the inclusion of the so-called cosmological constant the accelerated expanding Universe can be described only by geometrical properties involved in $f(Q)$ gravity. In this connection it is to be noted that the comprehending the observational data (such as the data of the Planck satellite \cite{Planck:2018vyg}) of the accelerated expanding Universe from the theoretical perspective becomes the first significant dispute to the GR. It has been shown in some research works \cite{Cai:2015emx,Jim2019a,BeltranJimenez:2017tkd,Solanki:2021qni} that various models incorporating $f(Q)$ gravity formulation matches observational results effectively and yields an alternative description to the dark energy in terms of gravitational interaction governed by the non-metricity scalar. Nashed et al. \cite{Nashed:2025rbf,Nashed:2025zpw} explored a few important implications of non linear functional form $f(Q)$ being quadratic in $Q$ to obtain an exact  solution representing physical features of astrophysical objects. In general, the alternative gravity theories become important tool to comprehend theoretically the disputes originating from astronomical data based of the gravitational dynamics. 

Now, some gravitational theories employ the technique of coupling among the geometrical quantities (or among the geometrical and matter quantities). For example, the non-metricity scalar is coupled with the trace of energy-momentum tensor ($T$) non-minimally to formulate another extension to the $f(Q)$ gravity. This class of gravity theory regarded as $f(Q,T)$ gravity \cite{Xu:2019sbp} or in the popular terminology {\it extended symmetric teleparallel gravity} which becomes interesting due to the matter-geometry non-minimal coupling leading to the transfer of energy-momentum between the geometry and matter. In this case, the functional form of the gravitational Lagrangian in the action is represented by a function of both $Q$ and $T$, i.e., $f(Q,T)$. 

Many researchers have studied various functional forms of $f(Q,T)$ as available in the literature  \cite{Arora:2020iva,Arora:2020tuk,Arora:2020met}. The functional form of the $f(Q,T)$ can be either linear or non-linear in  $Q$ and $T$ depending on the choices. In this connection the  $f(R,T)$ gravity theory proposed by Harko et al. \cite{Harko:2011kv}  is worth mentioning where $T$ and $R$ are coupled together. Various functional forms of $R$ and $T$ in $f(R,T)$ gravity model provide many research studies focusing mainly on the thermodynamical \cite{Momeni:2015fyt,Hazarika:2024cji}, astrophysical systems \cite{Sarkar:2024bxk,Lu:2024qne,Errehymy:2024spg,Malik:2024boe,Malik:2024wbj} and cosmological events \cite{Singh:2024kez,Fortunato:2023ypc,Malik:2024dwd,Koussour:2024glo}. So, this innovative approach of coupling geometric quantity with $T$ offers  aim to deepen insight into the Universe’s fundamental dynamics as well as with particular emphasis on the characteristics of compact objects. The configuration of compact stars have been studied in Tolman–Kuchowicz spacetime utilizing the coupling between Gauss-Bonnet term and $T$ in a gravitational theory known as $f(G,T)$ gravity  \cite{Naz:2023nmd}. Galaxies are built from astrophysical bodies such as compact objects that act as their core constituents and are embedded into the large-scale structure of the cosmic web. When gravity’s inward pull drives a star to collapse, it generates new compact objects. To unravel their internal structure of the compact objects it is essential to obtain exact analytical solutions to the nonlinear field equations in the framework of any gravity theory. Although these equations are highly nonlinear partial differential equations,  researchers have successfully  obtained exact and physically valid solutions to the field equations in the various  gravitational models in the area of both astrophysics and cosmology. The  first such exact solution obtained by Schwarzschild to describe a spherically symmetric astrophysical body in a vacuum \cite{Schwarzschild:1916uq} under the framework of GR. 

Compact stars considered to be as self gravitating stellar systems having compactness within the Buchdahl limit \cite{Buchdahl:1959zz} in the case of the isotropic pressures. However, this limit is subject to be modified in the presence of anisotropy within the self gravitating configurations. The anisotropy can occur due to the diverse scenario  such as the highly dense regions \cite{Ruderman:1972aj,Canuto:1974gi}, superfluidity \cite{Carter:1998rn}, pion condensation and solid cores \cite{Sawyer:1973fv} etc. Primary research studies based on the anisotropy in the compact stars in the background of GR explore various solutions in static and spherically symmetric spacetime. However, finding an anisotropic solution to the highly nonlinear gravitational field equations appears to be a difficult task when the number of unknown physical quantities exceeds the number of field equations. So, different techniques have been employed by the scientists to ease this task and to support the development of realistic solutions. 

Astrophysical objects such as neutron stars have the matter density at its core region of the order of super-nuclear density which is of highly dense matter as observed in the Universe. Eventually, neutron stars emerge as a significant area of research regarding the relations between the density and pressures in its dense core. According to the observational results obtained by NICER (Neutron Star Interior Composition Explorer), the predicted radii of the massive pulsar PSR J0740+6620 are  $R = 13.7^{+2.6}_{-1.5}$ km for the mass $2.08 \pm 0.07$ $M_{\odot}$ \cite{Miller:2021qha} and $R = 12.39^{+1.30}_{-0.98}$ km for the mass  $2.072^{+0.067}_{-0.066}$ $M_{\odot}$ \cite{Riley:2021pdl}. One can note that the mass-radius  measurements show significant impacts to put limitations on the equation of state (EOS) of the dense matter. Now, consider the maximum mass ($M_{max}$) of neutron star in the case of PSR J0952–0607 which is a companion in the black widow system measured to be  $2.35 \pm 0.17$ $M_{\odot}$  surpassing the previous limits on $M_{max}$ subject to PSR J2215–5135 \cite{Ravi:2017npf,Patra:2023jvv}. 

Observations of pulsars above 2 $M_{\odot}$ signify the need of stiffer EOS to stabilize the highly dense neutron stars. In spite of displaying almost similar radii, NICER’s measurements of PSR J0740+6620 whose mass is larger than that of PSR J0030+0451 pose a considerable challenge to the gravitational models predicting extremely compressible highly dense matter. Significant progress has been made in inferring phases of neutron star's dense matter supported by the rapid growth of neutron star  observations as well as theoretical calculations. In this regard, gravitational wave (GW) measurements play a leading role to understand the properties of neutron star at recent times.

GW has revolutionized our perspective to the Universe by furnishing pioneering revelations into the astrophysical as well as  cosmological events.  A new way of understanding the enigmas of the cosmos have been initiated post-observation of the LIGO-Virgo-KAGRA network. Particularly, merger events like GW170817 offers observational data of GW which came from the late inspiral of binary system of neutron star which could constrain the tidal deformability factor and equation of state. Further, the electromagnetic counterpart to the GW170817 revealed that the structure of a highly massive neutron star turned into a black hole due to gravitational collapse about the estimated time of one second after post-merger \cite{Gill:2019bvq,Murguia-Berthier:2020tfs}. Notably, mergers of  other binary systems such as black hole binary and  neutron star-black hole binary can rise to the GW signals. It is to be noted that the initial observational results of merger GW170817 set the upper bound of neutron star to be $2.17$ $M_{\odot}$ \cite{Margalit:2017dij}.  Later, the $M_{max}$ of neutron star predicted to be within the range 2.15--2.25 $M_{\odot}$ \cite{Ruiz:2017due,Rezzolla:2017aly} as implied by the missing relativistic and optical counterpart in merger GW170817. On the other hand, even higher values of $M_{max}$ between 2.5--2.65  $M_{\odot}$ \cite{LIGOScientific:2020zkf} has been observed in another gravitational merger event, namely GW190814. This range of observed  mass of GW190814 lies typically in the mass gap region. More accurate and sensitive GW signals with high values of signal-to-noise ratio are expected to obtain after the merger of the highly massive neutron stars by the third generation GW observatories \cite{Punturo:2010zza,Evans:2021gyd}. 

A non-singular gravitational model with anisotropic configuration have been developed in a recent study under the framework of $f(Q,T)$ gravity by utilizing observational constraints of the  pulsar SAX J1748.9-2021 \cite{Sharif:2025vvd}. Another solution for the anisotropic fluid configuration have been derived by Nashed el al. \cite{Nashed:2024pjc} associated with the observational data of SAX J1748.9-2021. Moreover, a recent investigation \cite{ZeeshanGul:2025wmz} considered charged as well as  anisotropic matter to develop gravitational model by employing the technique of extended gravitational decoupling  in the background  $f(Q,T)$ gravity. Further, the vanishing complexity method have been applied successfully to generate a solution to the modified field equations in  $f(Q,T)$ gravity \cite{Samanta:2025ueo}. 

Bhagat et al. \cite{Bhagat:2025bdk} have presented a few physical outcomes of the gravitational model in $f(Q,T)$ gravity utilizing  Markov chain Monte Carlo (MCMC) method and Bayesian information criterion (BIC). These outcomes in  $f(Q,T)$ gravity basically emulate $\Lambda$CDM model with another plausible explanation in the field of cosmology. The characteristic behavior of spatial homogeneity and anisotropy of the Universe have been analyzed in an investigation \cite{Zhadyranova:2024lyz} considering  Bianchi type I  line element in $f(Q,T)$ gravity. One can find some other investigations \cite{Narawade:2023pze,Najera:2021afa,Malik:2025ntj,Nashed:2025usa} which analyze the implications of $f(Q,T)$ gravity in the field of cosmology. 
 
Hence, inspired from the aforementioned theories and facts we would like to study physical properties of highly dense stars with anisotropic matter distributions in the framework of $f(Q,T)$ gravity. In addition to this we want to investigate the mass-radius relationship  and to explore upper bound of $M_{max}$ of neutron star in the present investigation. So, as an outline in the present article we introduce the basic theory of extended symmetric teleparallel gravity (i.e., $f(Q,T)$  gravity) briefly and present the gravitational modified field equations in Section \ref{II}. Then we find the anisotropic solutions to the modified field equations in the background of $f(Q,T)$ gravity in Section \ref{III}. Thereafter in Section \ref{IV}, we obtain the expressions for various unknown constants in the present model by using boundary conditions. The physical analysis and the final remarks are presented in Section \ref{V} and \ref{VI}, respectively.

\section{A brief Overview of Extended Symmetric Teleparallel Gravity And Corresponding Field Equations For Anisotropic Matter Distributions}\label{II}
We start with modified action integral in $f(Q,T)$ gravity as \cite{Xu:2019sbp}
\begin{eqnarray}
    \label{1}
&&\hspace{-0.5cm} S=\int \sqrt{-g}\Big[\frac{1}{2 }f(Q,T)+\mathcal{L}_M\Big] d^4x.
\end{eqnarray}

The Lagrangian density for matter is represented by $\mathcal{L}_M$, and it is connected with the energy-momentum tensor $T_{\alpha \eta}$. The determinant of the metric tensor is represented by $g$, whereas the gravitational sector is symbolized as $\Theta$. For the sake of simplicity, we follow the convention \(8\pi G = c = 1\) throughout this study.

The non-metricity tensor associated with the affine connection can be stated as: 
\begin{eqnarray}
\label{4}
Q_{\mu\alpha \eta}\equiv \nabla_\mu g_{\alpha \eta}=\partial_{\mu}\,g_{\alpha \eta}-\Gamma^{\nu}{}_{\mu\alpha}\,g_{\nu  \eta}-\Gamma^{\nu}{}_{\mu \eta}\,g_{\alpha\nu},
\end{eqnarray}
where the usual affine connection is represented by $\Gamma^{\nu}{}_{\mu\alpha}$ and is expressed as
\begin{eqnarray}
    \Gamma^{\nu}{}_{\mu\alpha} = L^{\nu}{}_{\mu\alpha}+K^{\nu}{}_{\mu \alpha}+\left\{
\begin{array}{c}
\nu \\ \mu ~\alpha 
\end{array}
\right\}.
\end{eqnarray}

The contorsion tensor, deformation tensor, Levi-Civita connection are represented as $L^{\nu}{}_{\mu\alpha}$, $K^{\nu}{}_{\mu \alpha}$, and $\left\{ \begin{array}{c} \nu \\ \mu ~\alpha \end{array} \right\}$, respectively. They can be expressed mathematically in their full expressions as follows:
\begin{eqnarray}
L^{\nu}{}_{\mu\alpha} = \frac{1}{2} Q^{\nu}{}_{\mu\,\alpha}-Q_{(\mu}{\,}^{\nu}{\,}_{\alpha)},\\
K^{\nu}{}_{\mu \alpha} = \frac{1}{2} T^{\nu}{}_{\mu \alpha} + T_{(\mu}{\,}^{\nu}{\,}_{\alpha)},\\
 \left\{
\begin{array}{c}
\nu \\ \mu \alpha 
\end{array}
\right\} = \frac{1}{2}\,g^{\nu\varphi}\left(\partial_{\mu}\, g_{\varphi\alpha}+\partial_{\alpha}\, g_{\nu \eta}-\partial_{\nu}\, g_{\mu\alpha}\right),
\end{eqnarray}
where $T^{\nu}{}_{\mu \alpha}$ denotes the torsion tensor. 

In the above, the superpotential $P_{\,\,\,\,\alpha \eta}^{\varphi}$, a crucial component of the STEGR formalism, is stated as follows: 
\begin{eqnarray}
\label{5}
P_{\,\,\,\,\alpha \eta}^{\varphi}=-\frac{1}{4}\, \nu^{\varphi}_{\,\,(\alpha\,} Q_{ \eta)}-\frac{1}{2}\, L^\varphi_{\,\,\,\,\alpha \eta}-\frac{1}{4}\left(\tilde{Q}^{\varphi}-Q^{\varphi}\right)g_{\alpha \eta},
\end{eqnarray}
and the contracted forms of non-metricity tensor are shown as follows:
\begin{equation*}
\tilde{Q}_{\varphi}=Q^{\alpha}_{\,\,\,\,\varphi\alpha} \quad \quad Q_{\varphi}=Q_{\varphi\,\,\,\,\,\alpha}^{\,\,\,\alpha}\, .
\end{equation*}

The non-metricity scalar ($Q$) is finally presented as
\begin{eqnarray}
\label{6}
Q=-Q_{\varphi\alpha \eta}P^{\varphi\alpha \eta}.
\end{eqnarray}

The $f(Q,T)$ gravity field equations are derived by applying the variational principle to the action \eqref{1}, changing it with regard to metric tensor $g^{\alpha \eta}$ as
\begin{eqnarray}
\label{field}
\frac{2}{\sqrt{-g}}\nabla_{\varphi}\left(f_{Q}\sqrt{-g}\,P^{\varphi}_{\,\,\,\,\alpha \eta}\right)+f_{Q}\left(P_{\alpha\varphi\mu}Q_{ \eta}^{\,\,\,\varphi\mu}-2Q^{\varphi\mu}_{\,\,\,\,\,\,\,\,\alpha}\, P_{\varphi\mu \eta}\right)+\frac{1}{2}f\,g_{\alpha \eta}=- T_{\alpha \eta}  \nonumber\\+f_{T} \left(T_{\alpha \eta}+\Phi_{\alpha \eta}\right),
\end{eqnarray}
where $f_Q = \frac{\partial f(Q,T)}{\partial Q}$, $f_T = \frac{\partial f(Q,T)}{\partial T}$. 

The usual way to write the energy-momentum tensor $T_{\alpha \eta}$ is as
\begin{eqnarray}
 T_{\alpha \eta} &=& -\frac{2}{\sqrt{-g}}\frac{\delta (\sqrt{-g}\mathcal{L}_m)}{\delta g^{\alpha \eta}}.
\end{eqnarray}

The hyper-momentum tensor, an extra unknown variable in Eq.~(\ref{field}), is denoted as $\Phi_{\alpha \eta}$ and is defined by the equation $\Phi_{\alpha \eta} = g^{\varphi  \eta}\frac{\delta T_{\varphi  \eta}}{\delta g^{\alpha \eta}}$. 

Furthermore, the further constraint may be obtained using Eq.~(\ref{1}) and can be interpreted as follows: 
\begin{eqnarray}
    \nabla^{\alpha}\nabla^{ \eta}(\sqrt{-g}\,f_{Q}P^{\varphi}{}_{\alpha \eta})=0.
\end{eqnarray}

The affine connection satisfies the following requirements of being curvature-free and torsion-free \cite{BeltranJimenez:2017tkd} as
\begin{eqnarray}\label{eq:13}
    \Gamma^{\varphi}{}_{\alpha \eta}=\left(\frac{\partial x^{\varphi}}{\partial \xi^{\alpha}}\right) \partial_{\alpha}\partial_{ \eta} \xi^{\alpha}.
\end{eqnarray}

Selecting a particular coordinate structure, referred to as the co-incident gauge, guarantees that the affine connection fulfills the condition $\Gamma^{\varphi}{}_{\alpha \eta}=0$. As a result, the non-metricity formula is simplified to
\begin{eqnarray}
    Q_{\mu\alpha \eta}\equiv \nabla_\mu \, g_{\alpha \eta}=\partial_{\mu}\,g_{\alpha \eta}.
\end{eqnarray}

Here, it is noticed that the metric tensor is the only variable to define the non-metricity.  Our present study is focused on static metric structure with spherical symmetry to show the internal structure of compact astrophysical objects. Therefore, we consider the following line element  
\begin{eqnarray}\label{eq:metric}
    ds^2=-e^{\mathcal{U}(r)} dt^2+e^{\mathcal{W}(r)} dr^2+r^2(d\theta^2+\sin^{2}\theta \, d\phi^2),
\end{eqnarray}
where $\mathcal{U}(r)$, and $\mathcal{W}(r)$ represent the space-time coordinates, metric potentials over the time, and radial directions, correspondingly.  Then, under the metric (\ref{eq:metric}), the non-metricity scalar can be expressed as
\begin{eqnarray}
    Q = - \frac{2 e^{-\mathcal{W}(r)}[r \,\mathcal{U}'(r)+1]}{r^2}.
\end{eqnarray}

In the present study, we are going to model the compact stars with an anisotropic fluid. In this regard, the the energy-momentum tensor can be written as: 
\begin{eqnarray}
 T_{\alpha \nu} = \rho^{\text{eff}}\, \eta_{\alpha } \eta_{\nu}  + p_t^{\text{eff}} (\eta_{\alpha } \eta_{\nu} +g_{ \alpha \nu }-\psi_{\alpha}\,\psi_{\nu})+p_r^{\text{eff}} \psi_{\alpha}\,\psi_{\nu}.  \label{em}
\end{eqnarray}

 In this case, $\rho^{\text{eff}}$ stands for matter density, $p_r^{\text{eff}}$ for radial pressure, and $p_t^{\text{eff}}$ for tangential pressure. The non-vanishing components of $T_{\alpha \eta}$ are: $(-\rho^{\text{eff}}, p_r^{\text{eff}}, p_t^{\text{eff}}, p_t^{\text{eff}})$. On the other hand, the Lagrangian for matter is set to $L_{m}=-\mathcal{P}=-\frac{p_r^{\text{eff}}+2p_t^{\text{eff}}}{3}$. To find each component of $\Phi_{\alpha \eta}$, use the formula $\Phi_{\alpha \eta}=g_{\alpha \eta}\mathcal{P}-2T_{\alpha \eta}$.
 
Under the following above assumption, we find three independent equations in extended symmetric teleparallel gravity for anisotropic matter distribution for the line \eqref{eq:metric} as
\begin{eqnarray}
    \label{FE1a}
&& \hspace{-0.2cm}8\pi \rho^{\text{eff}}= \frac{1}{2r^2e^{\mathcal{W}}} \Big[2 r f_{QQ} Q^\prime (e^{\mathcal{W}}-1) + f_Q \big((e^\mathcal{W}-1) (2+r \mathcal{U}^\prime)+(1+e^\mathcal{W}) r \mathcal{W}^\prime \big)+f r^2 e^{\mathcal{W}} \Big] \nonumber\\
&& \hspace{11.8cm}-f_{T} (\mathcal{P}+\rho^{\text{eff}}),~~~~~~~~~~\\
\label{FE2a}
&& \hspace{-0.2cm}8\pi p_r^{\text{eff}}= -\frac{1}{2r^2e^{\mathcal{W}}} \Big[2 r f_{QQ} Q^\prime (e^{\mathcal{W}}-1) + f_Q \big((e^\mathcal{W}-1) (2+r \mathcal{U}^\prime+r \mathcal{W}^\prime)- 2r \mathcal{U}^\prime +f r^2 e^{\mathcal{W}} \big)\nonumber\\
&& \hspace{9.5cm}+f r^2 e^{\mathcal{W}} \Big]+f_{T} (\mathcal{P}-p_r^{\text{eff}}) ,~~~~\\ 
 \label{FE3a}
&& \hspace{-0.2cm}8\pi p_t^{\text{eff}}=- \frac{1}{4 r e^{\mathcal{W}}} \Big[-2 r f_{QQ} Q^\prime \mathcal{U}^\prime + f_Q \big(2\mathcal{U}^\prime (e^\mathcal{W}-2) -r \mathcal{U}^{\prime 2} +\mathcal{W}^\prime (2e^{\prime}+r \mathcal{U}^{\prime}) -2 r \mathcal{U}^{\prime \prime}\big)\nonumber\\
&& \hspace{9.5cm}+2 f r e^{\mathcal{W}}\Big] +f_{T} (\mathcal{P}+p_t^{\text{eff}}).~~~~~~~
\end{eqnarray}

 We shall now examine a certain functional form of gravity, which is represented by the equation $f(Q)=\psi_1 Q +\psi_2 T$. As a result, the field equations (\ref{FE1a})--(\ref{FE3a}) may be rewritten in the following manner:
\begin{eqnarray}
    \label{FE1}
&& \hspace{-1cm} \frac{8\pi \rho^{\text{eff}}}{\psi_1} + \frac{\psi_2}{3\psi_1} (3\rho^{\text{eff}}+p_r^{\text{eff}}+2p_t^{\text{eff}})-\frac{\psi_2}{2\psi_1}(\rho^{\text{eff}}-p_r^{\text{eff}}-2p_t^{\text{eff}}) = \Big[e^{-\mathcal{W}}\Big(\frac{\mathcal{W}^{\prime}}{r}-\frac{1}{r}\Big)+\frac{1}{r^2}\Big],\\
\label{FE2}
&& \hspace{-1cm} \frac{8\pi p_r^{\text{eff}}}{\psi_1} - \frac{2\psi_2}{3\psi_1} (p_t^{\text{eff}}-p_r^{\text{eff}})+\frac{\psi_2}{2\psi_1}(\rho^{\text{eff}}-p_r^{\text{eff}}-2p_t^{\text{eff}}) = \Big[e^{-\mathcal{W}}\Big(\frac{\mathcal{U}^{\prime}}{r}+\frac{1}{r^2}\Big)-\frac{1}{r^2}\Big],~~\\ 
 \label{FE3}
&& \hspace{-1cm} \frac{8\pi p_t^{\text{eff}}}{\psi_1} - \frac{\psi_2}{3\psi_1} (p_r^{\text{eff}}-p_t^{\text{eff}})+\frac{\psi_2}{2\psi_1}(\rho^{\text{eff}}-p_r^{\text{eff}}-2p_t^{\text{eff}}) = \Big[e^{-\mathcal{W}}\Big\{\frac{\mathcal{U}^{\prime \prime}}{2}+\Big(\frac{\mathcal{U}^{\prime}}{4}+\frac{1}{2 r}\Big) \nonumber \\
&& \hspace{10cm} \left(\mathcal{U}^{\prime}-\mathcal{W}^{\prime}\right)\Big\}\Big],~~~~~~~
\end{eqnarray}
and the associated TOV equation may be stated as follows:
\begin{eqnarray}
&&\hspace{-0.5cm}   - \frac{\mathcal{U}^\prime}{2} (\rho^{\text{eff}}+p_r^{\text{eff}})-\frac{dp_r^{\text{eff}}}{dr}+\frac{2}{r}(p_t^{\text{eff}}-p_r^{\text{eff}})-\frac{\psi_2}{6(8\pi+\psi_2)} \Big[3\frac{d\rho^{\text{eff}}}{dr}-5 \frac{dp_r^{\text{eff}}}{dr}-10 \frac{dp_t^{\text{eff}}}{dr}\Big]=0.~~~~~~~~\label{eq21}
\end{eqnarray}

To asses the internal structure of the compact stars, we must find the explicit expressions for $\rho^{\text{eff}}$, $p_r^{\text{eff}}$, and $p_t^{\text{eff}}$. After solving the Eqs. (\ref{FE1})--(\ref{FE3}), we obtain the explicit equations for $\rho^{\text{eff}}$, $p_r^{\text{eff}}$, and $p_t^{\text{eff}}$ as
\begin{eqnarray}
  &&\hspace{-0.5cm}  \rho^{\text{eff}}=-\frac{\psi _1}{24 r^2 \left(4 \pi -\psi _2\right) \left(\psi _2+8 \pi \right)} \Big[\psi _2 \big(e^{-\mathcal{W}} \left(10\, \mathcal{U}'' r^2+5 \mathcal{U}'^2 r^2+20 \mathcal{U}' r-8\right)-\mathcal{W}' e^{-\mathcal{W}} r (5 \mathcal{U}' r-8)\nonumber\\
  &&\hspace{0.5cm}+8\big)+96 \pi  (-\mathcal{W}' e^{-\mathcal{W}} r+e^{-\mathcal{W}}-1)\Big],\label{eq2.24}\\
   &&\hspace{-0.5cm}  p_r^{\text{eff}}=\frac{\psi _1}{24 r^2 \left(4 \pi -\psi _2\right) \left(\psi _2+8 \pi \right)} \Big[\psi _2 \big(e^{-\mathcal{W}} \left(10 \mathcal{U}'' r^2+5 \mathcal{U}'^2 r^2-4 \mathcal{U}' r-8\right)-\mathcal{W}' e^{-\mathcal{W}} r (5 \mathcal{U}' r+16)\nonumber\\
   &&\hspace{0.5cm}+8\big)+96 \pi  (\mathcal{U}' e^{-\mathcal{W}} r+e^{-\mathcal{W}}-1)\Big],\label{eq2.25}\\
   &&\hspace{-0.5cm}  p_t^{\text{eff}}=\frac{\psi _1}{24 r^2 \left(4 \pi -\psi _2\right) \left(\psi _2+8 \pi \right)} \Big[24 \pi  r \left(e^{-\mathcal{W}} \left(2 \mathcal{U}'' r+\mathcal{U}'^2 r+2 \mathcal{U}'\right)-\mathcal{W}' e^{-\mathcal{W}} (\mathcal{U}' r+2)\right)-\psi _2 \nonumber\\
   &&\hspace{0.5cm} \times \left\{e^{-\mathcal{W}} \left(2 \mathcal{U}'' r^2+\mathcal{U}'^2 r^2-8 \mathcal{U}' r-16\right)-\mathcal{W}' e^{-\mathcal{W}} r (\mathcal{U}' r-4)+16\right\}\Big]. \label{eq2.26}
\end{eqnarray}

\section{Most General Exact Anisotropic Solutions in Extended Symmetric Teleparallel Gravity}\label{III}

 Our main objective is to obtain the most general exact solution of the field equations (\ref{FE1})--(\ref{FE3}) that gives a well-behaved model. To do this, we must subtract Eq. (\ref{FE2}) from Eq. (\ref{FE3}) for the purpose of getting the anisotropy condition in $f(Q,T)$-gravity as 
\begin{eqnarray}
   &&\hspace{-0.5cm} \Big[e^{-\mathcal{W}}\Big\{\frac{\mathcal{U}^{\prime \prime}}{2}+\Big(\frac{\mathcal{U}^{\prime}}{4}+\frac{1}{2 r}\Big) \left(\mathcal{U}^{\prime}-\mathcal{W}^{\prime}\right)-\Big(\frac{\mathcal{U}^{\prime}}{r}+\frac{1}{r^2}\Big)\Big\} +\frac{1}{r^2}\Big]=\frac{(8\pi+\psi_2) (p_t^{\text{eff}}-p_r^{\text{eff}})}{\psi_1}. \label{master1} 
\end{eqnarray}

The above anisotropy condition represents the most general differential equation in $f(Q,T)$ gravity. The Eq. (\ref{master1}) is highly non-linear, and finding its exact solution is not always an easy task. Therefore, first we utilize the transformation $\mathcal{A}=\mathcal{X} r^2$,~~ $ e^{\mathcal{U}}= \mathcal{H}^2$ and $e^{-\mathcal{W}}=\mathcal{N}$ in Eq. (\ref{master1}) to express the master equation in standard form
\begin{eqnarray}
    4\mathcal{A}^2 \mathcal{N} \frac{d^2 \mathcal{H}}{d\mathcal{A}^2}+2\mathcal{A}^2 \frac{d \mathcal{H}}{d\mathcal{A}} \frac{d \mathcal{N}}{d\mathcal{A}}  + \Big( \mathcal{A} \frac{d \mathcal{N}}{d\mathcal{A}} - \mathcal{N}+1-\frac{ \mathcal{A} (8\pi+\psi_2) \nu }{\mathcal{X} \psi_1} \Big)\mathcal{H} =0.~~~~\label{master2}
\end{eqnarray}

Then final form of expressions for $\rho^{\text{eff}}$, $p_r^{\text{eff}}$ and $p_t^{\text{eff}}$ in terms of $\mathcal{A}$, $\mathcal{H}$, and $\mathcal{N}$ are calculated from Eqs.~(\ref{eq2.24})--(\ref{eq2.26}) as: 
\begin{small}
\begin{eqnarray}
 &&\hspace{-0.5cm}  \rho^{\text{eff}}=-\frac{\psi _1 \left\{\psi _2 \left(10 \mathcal{\overline{\overline{H}}} \mathcal{A}^2 \mathcal{N}+5 \mathcal{\overline{H}} \mathcal{A} (\mathcal{\overline{N}} \mathcal{A}+3 \mathcal{N})-2 \mathcal{\overline{N}} \mathcal{A} \mathcal{H}-\mathcal{H} \mathcal{N}+\mathcal{H}\right)+12 \pi  \mathcal{H} (2 \mathcal{\overline{N}} \mathcal{A}+\mathcal{N}-1)\right\}}{3 \mathcal{A} \left(4 \pi -\psi _2\right) \left(\psi _2+8 \pi \right) \mathcal{H}},~~~~\label{eq27}\\
    &&\hspace{-0.5cm} p_r^{\text{eff}}= \frac{\psi _1 \left\{\psi _2 \left(10 \mathcal{\overline{\overline{H}}} \mathcal{A}^2 \mathcal{N}+\mathcal{\overline{H}} \mathcal{A} (5 \mathcal{\overline{N}} \mathcal{A}+3 \mathcal{N})+4 \mathcal{\overline{N}} \mathcal{A} \mathcal{H}-\mathcal{H} \mathcal{N}+\mathcal{H}\right)+12 \pi  (4 \mathcal{\overline{H}} \mathcal{A} \mathcal{N}+\mathcal{H} (\mathcal{N}-1))\right\}}{3 \mathcal{A} \left(4 \pi -\psi _2\right) \left(\psi _2+8 \pi \right) \mathcal{H}},\label{eq28}\\
  &&\hspace{-0.5cm} p_t^{\text{eff}}= \frac{\psi _1 \left\{\psi _2 \left(-2 \mathcal{\overline{\overline{H}}} \mathcal{A}^2 \mathcal{N}+\mathcal{\overline{H}} \mathcal{A} (3 \mathcal{N}-\mathcal{\overline{N}} \mathcal{A})+\mathcal{H} (\mathcal{\overline{N}} \mathcal{A}+2 \mathcal{N}-2)\right)+12 \pi  \mathcal{A} (4 \mathcal{\overline{\overline{H}}} \mathcal{A} \mathcal{N}+2 \mathcal{\overline{H}} \mathcal{\overline{N}} \mathcal{A}+P_{t1})\right\}}{3 \mathcal{A} \left(4 \pi -\psi _2\right) \left(\psi _2+8 \pi \right) \mathcal{H}},~~~~~~~~
\end{eqnarray}
\end{small}
in which setting, $P_{t1}=4 \mathcal{\overline{H}} \mathcal{N}+\mathcal{\overline{N}} \mathcal{H}$, $\Delta^{\text{eff}}=p_t^{\text{eff}}-p_r^{\text{eff}}$, $\mathcal{\overline{H}}=\frac{d\mathcal{H}}{d\mathcal{A}}$, $\mathcal{\overline{\overline{H}}}=\frac{d^2\mathcal{H}}{d\mathcal{A}^2}$ and $\mathcal{\overline{N}}=\frac{d\mathcal{N}}{d\mathcal{A}}$. The master equation (\ref{master2}) has three unknowns: $\mathcal{H}$, $\mathcal{N}$, and $\Delta^{\text{eff}}$. Consequently, we use a physically feasible equation for $\mathcal{N}$ as follows:
\begin{eqnarray}
   \mathcal{N}={\frac{7 (\mathcal{A}+1)^2-8 \mathcal{A}^2-24\mathcal{A}}{7 (1+2\mathcal{A}+\mathcal{A}^2)}}. \label{eq29}
\end{eqnarray}

Upon substituting Eq. (\ref{eq29}) into Eq. (\ref{master2}), we get   
\begin{eqnarray}
  &&\hspace{-0.0cm}   \frac{d^2\mathcal{H}}{d\mathcal{A}^2} - \frac{4  (\mathcal{A}-3) \mathcal{A} }{(\mathcal{A}+1) \left(\mathcal{A}^2+10 \mathcal{A}-7\right)} \frac{d\mathcal{H}}{d\mathcal{A}}  +\frac{2(1+\mathcal{A})^2}{\mathcal{A} (7-10\mathcal{A}-\mathcal{A}^2)} \bigg[ \frac{\mathcal{A}(\mathcal{A}+5)}{(1+\mathcal{A})^3}-\frac{7 (8\pi+\psi_2) \Delta}{8\mathcal{X} \psi_1}\bigg] \mathcal{H}=0. ~~~~~~~~~\label{eq30} 
\end{eqnarray}

 It is clear that the previously described Eq. (\ref{eq30}) necessitates a suitable formula for the anisotropy $\Delta^{\text{eff}}$ to enable the complete integration of Eq.~(\ref{eq30}). It is observed that for $\Delta^{\text{eff}}=0$, $\mathcal{H}=(1+\mathcal{A})^2$ constitutes a specific solution of Eq.~(\ref{eq30}). But here we are looking for a non-vanishing expression for anisotropy. To find this, we need to adjust this specific solution by including an anisotropy parameter $\beta$ and assume ${\mathcal{H}}=(1+\beta+\mathcal{A})^2$ is a specific solution. Based on this hypothesis, we select a non-zero expression for the anisotropy $\Delta^{\text{eff}}$ that looks like as:    
\begin{eqnarray}
\Delta^{\text{eff}} =\frac{8  \psi_1 \mathcal{X} \mathcal{A} \beta  \left[2 \mathcal{A} (\mathcal{A}+8)+\beta \mathcal{A} +5 \beta -2\right]}{7 (8\pi+\psi_2)(\mathcal{A}+1)^3 (\mathcal{A}+\beta +1)^2},~~~~\label{eq31}
\end{eqnarray}
where $\beta$ represents a positive constant.  

To maintain positive anisotropy across the stellar configuration, the expression $$\left[2 \mathcal{A} (\mathcal{A}+8)
+\beta \mathcal{A} +5 \beta -2\right]$$ must be greater than zero and this requirement imposes the restriction as follows: 
\begin{eqnarray}
    0 \le \mathcal{A}=\mathcal{X} r^2 < \frac{-(16+\beta) + \sqrt{\beta^2-8\beta+272}}{4}. ~~~~~~~~~\label{eq31a} 
\end{eqnarray}

We derive the final version of the master equation as follows after inserting the aforementioned expression (\ref{eq31}) into Eq. (\ref{eq30}) as,  
\begin{eqnarray}
&&\hspace{-0.2cm}   \frac{d^2\mathcal{H}}{d\mathcal{A}^2}-\frac{4 (\mathcal{A}-3) }{[\mathcal{A} (\mathcal{A}+10)-7] (\mathcal{A}+1) }\frac{d\mathcal{H}}{d\mathcal{A}}+\frac{8 (\mathcal{A}-3) \beta -2 (\mathcal{A}+5)  (\mathcal{A}+1)^2 }{[\mathcal{A} (\mathcal{A}+10)-7] (\mathcal{A}+1) (\mathcal{A}+\beta +1)^2}\mathcal{H}=0. ~~~~~~~~\label{eq32}
\end{eqnarray}

Upon examining Eq.~(\ref{eq32}), it turns out that ${\mathcal{H}}=(1+\beta+\mathcal{A})^2$ is one of the solutions to the aforementioned equation. Thus, this value of $\hat{\mathcal{H}}$ can be assigned as a particular solution of Eq. (\ref{eq32}). The complete solution can be derived through the change of dependent variable method by proposing the transformation $\mathcal{H}=\hat{\mathcal{H}}\, \mathcal{H}_0$, where $\hat{\mathcal{H}}=(1+\beta+\mathcal{A})^2$.

Again, $\mathcal{H}_0$ can be determined through the integration of Eq.~(\ref{eq32}) as: 
 \begin{eqnarray}
 &&\hspace{-0.2cm} \mathcal{H}_0= \frac{1}{3 (\beta +1)^3} \Bigg[3 \mathcal{C}_1-\frac{\mathcal{C}_2}{\left(\beta ^2-8 \beta -16\right)^2}  \Bigg(\frac{\sqrt{7-\mathcal{A} (\mathcal{A}+10)}}{[(\beta -8) \beta -16] (\mathcal{A}+\beta +1)^3} \Big[3 (\mathcal{A}-3) \beta ^4  +(\mathcal{A} (\mathcal{A}\nonumber\\
 && \hspace{0.5cm}+16)+119) \beta ^3+2 (\mathcal{A} (5 \mathcal{A}+86)-47) \beta ^2+16 (5 (\mathcal{A}-8) \mathcal{A}-53) \beta -96 (\mathcal{A}+1) (3 \mathcal{A}+7)+3 \beta ^5\Big] \nonumber\\
 && \hspace{0.5cm} +\frac{12 ((\beta -4) \beta  (\beta +8)+128)}{(16-(\beta -8) \beta )^{3/2}}\tanh ^{-1}\left(\frac{\mathcal{A} (\beta -4)+5 \beta +12}{\sqrt{7-\mathcal{A} (\mathcal{A}+10)} \sqrt{16-(\beta -8) \beta }}\right)\Bigg)\Bigg],~~~~~~~~
\end{eqnarray}
where the integration constants $\mathcal{C}_1$ and $\mathcal{C}_2$ are used. 

The general solution to Eq.~(\ref{eq29}) may therefore be written as
\begin{eqnarray}
e^{\mathcal{U}/2}&=&\mathcal{H}=\hat{\mathcal{H}}~\mathcal{H}_0 =\frac{(1 + \mathcal{A} + \beta)^2 }{3 (\beta +1)^3} \Bigg[-\frac{\mathcal{C}_2}{\left(\beta ^2-8 \beta -16\right)^2} \Bigg(\frac{\sqrt{7-\mathcal{A} (\mathcal{A}+10)}}{[(\beta -8) \beta -16] (\mathcal{A}+\beta +1)^3} \Big[3 (\mathcal{A}-3) \beta ^4 \nonumber\\&&\hspace{-0.5cm} +(\mathcal{A} (\mathcal{A}+16)+119) \beta ^3+2 (\mathcal{A} (5 \mathcal{A}+86)-47) \beta ^2+16 (5 (\mathcal{A}-8) \mathcal{A}-53) \beta -96 (\mathcal{A}+1) (3 \mathcal{A}+7)\nonumber\\&&\hspace{-0.5cm}+3 \beta ^5\Big] +\tanh ^{-1}\left(\frac{\mathcal{A} (\beta -4)+5 \beta +12}{\sqrt{7-\mathcal{A} (\mathcal{A}+10)} \sqrt{16-(\beta -8) \beta }}\right)\frac{12 ((\beta -4) \beta  (\beta +8)+128)}{(16-(\beta -8) \beta )^{3/2}}  \Bigg)+3 \mathcal{C}_1\Bigg].~~~~~~~\label{eq34}
\end{eqnarray}

After plugging of $\mathcal{H}$ and $\mathcal{N}$ in Eqs. (\ref{eq27})--(\ref{eq29}), we deriv the final expressions for $\rho^{\text{eff}}$, $p_r^{\text{eff}}$ and $p_t^{\text{eff}}$ as
\begin{eqnarray}
  &&\hspace{-0.2cm}  \rho^{\text{eff}}=\frac{\psi _1 \mathcal{X}}{21 (\mathcal{A}+1)^3 \left(4 \pi -\psi _2\right) \left(\psi _2+8 \pi \right) \mathcal{H}} \Big[\psi _2 \Big(10 \mathcal{F}_2(\mathcal{A}) \mathcal{A} \left(\mathcal{A}^3+11 \mathcal{A}^2+3 \mathcal{A}-7\right)+5 \mathcal{F}_1(\mathcal{A}) \nonumber\\
  &&\hspace{0.5cm} \times \big(3 \mathcal{A}^3+25 \mathcal{A}^2+33 \mathcal{A}-21\big)-8 \left(\mathcal{A}^2+2 \mathcal{A}+9\right) \mathcal{H}\Big)+96 \pi  \left(\mathcal{A}^2+2 \mathcal{A}+9\right) \mathcal{H}\Big],~~~~~~~~\\
   &&\hspace{-0.2cm}  p_r^{\text{eff}}= -\frac{\psi _1 \mathcal{X}}{21 (\mathcal{A}+1)^3 \left(4 \pi -\psi _2\right) \left(\psi _2+8 \pi \right) \mathcal{H}} \Big[\psi _2 \Big(10 \mathcal{F}_2(\mathcal{A}) \mathcal{A} \left(\mathcal{A}^3+11 \mathcal{A}^2+3 \mathcal{A}-7\right)+\mathcal{F}_1(\mathcal{A}) \nonumber\\
   && \hspace{0.5cm} \times \big(3 \mathcal{A}^3-7 \mathcal{A}^2+129 \mathcal{A}-21\big)-8 \left(\mathcal{A}^2+8 \mathcal{A}-9\right) \mathcal{H}\Big)+48 \pi  (\mathcal{A}+1) \big(\mathcal{F}_1(\mathcal{A}) \left(\mathcal{A}^2+10 \mathcal{A}-7\right)\nonumber\\
   && \hspace{0.5cm}+2 (\mathcal{A}+3) \mathcal{H}\big)\Big]\\
  &&\hspace{-0.2cm} p_t^{\text{eff}}=-\frac{\psi _1 \mathcal{X}}{21 (\mathcal{A}+1)^3 \left(4 \pi -\psi _2\right) \left(\psi _2+8 \pi \right) \mathcal{H}} \Big[\psi _2 \Big(-2 \mathcal{F}_2(\mathcal{A}) \mathcal{A} \left(\mathcal{A}^3+11 \mathcal{A}^2+3 \mathcal{A}-7\right)+\mathcal{F}_1(\mathcal{A}) \nonumber\\
  &&\hspace{0.5cm} \times \big(3 \mathcal{A}^3+41 \mathcal{A}^2-15 \mathcal{A}-21\big)+8 \left(2 \mathcal{A}^2+7 \mathcal{A}+9\right) \mathcal{H}\Big)+48 \pi  \Big(\mathcal{F}_2(\mathcal{A}) \mathcal{A} \left(\mathcal{A}^3+11 \mathcal{A}^2+3 \mathcal{A}-7\right)\nonumber\\
  &&\hspace{0.5cm}+\mathcal{F}_1(\mathcal{A}) \big(\mathcal{A}^3+7 \mathcal{A}^2+15 \mathcal{A}-7\big)-2 (\mathcal{A}-3) \mathcal{H}\Big)\Big],
\end{eqnarray}
where $\mathcal{H}$ is given by Eq.~(\ref{eq34}), while $\mathcal{F}_1(\mathcal{A})$ and $\mathcal{F}_2(\mathcal{A})$ are given below:
\begin{eqnarray}
 &&\hspace{-0.0cm}  \mathcal{F}_1(\mathcal{A})=\frac{\mathcal{C}_2 (\mathcal{A}+1)}{\sqrt{-\mathcal{A}^2-10 \mathcal{A}+7} (\beta +1)^3 (\mathcal{A}+\beta +1)^2}+\frac{2 (\mathcal{A}+\beta +1) }{3 (\beta +1)^3}\Bigg[3 \mathcal{C}_1-\frac{\mathcal{C}_2}{\left(\beta ^2-8 \beta -16\right)^2} \nonumber\\&& \hspace{1cm}\times \Bigg(\frac{\sqrt{7-\mathcal{A} (\mathcal{A}+10)}}{((\beta -8) \beta -16) (\mathcal{A}+\beta +1)^3}  \Big[3 (\mathcal{A}-3) \beta ^4+(\mathcal{A} (\mathcal{A}+16)+119) \beta ^3+2 (\mathcal{A} (5 \mathcal{A}+86)\nonumber\\&& \hspace{1cm}-47) \beta ^2+16 (5 (\mathcal{A}-8) \mathcal{A}-53) \beta -96 (\mathcal{A}+1) (3 \mathcal{A}+7)+3 \beta ^5\Big]+\frac{12 ((\beta -4) \beta  (\beta +8)+128) }{(16-(\beta -8) \beta )^{3/2}} \nonumber\\&& \hspace{1cm} \times \tanh ^{-1}\left(\frac{\mathcal{A} (\beta -4)+5 \beta +12}{\sqrt{7-\mathcal{A} (\mathcal{A}+10)} \sqrt{16-(\beta -8) \beta }}\right)\Bigg)\Bigg], \nonumber\\
   &&\hspace{-0.0cm} \mathcal{F}_2(\mathcal{A})=\frac{}{6 (\beta +1)^3}\Bigg[-\frac{4 \mathcal{C}_2}{\left(\beta ^2-8 \beta -16\right)^2} \Bigg(\frac{\sqrt{-\mathcal{A}^2-10 \mathcal{A}+7}}{((\beta -8) \beta -16) (\mathcal{A}+\beta +1)^3} \Big[3 (\mathcal{A}-3) \beta ^4+(\mathcal{A} (\mathcal{A}+16)\nonumber\\&& \hspace{1cm}+119) \beta ^3+2 (\mathcal{A} (5 \mathcal{A}+86)-47) \beta ^2+16 (5 (\mathcal{A}-8) \mathcal{A}-53) \beta -96 (\mathcal{A}+1) (3 \mathcal{A}+7)+3 \beta ^5\Big]\nonumber\\&& \hspace{1cm} +\frac{12 ((\beta -4) \beta  (\beta +8)+128) }{\left(-\beta ^2+8 \beta +16\right)^{3/2}}  \tanh ^{-1}\left(\frac{\mathcal{A} (\beta -4)+5 \beta +12}{\sqrt{7-\mathcal{A} (\mathcal{A}+10)} \sqrt{16-(\beta -8) \beta }}\right)\Bigg)\nonumber\\&& \hspace{1cm}+\frac{6 \mathcal{C}_2 (\mathcal{A}+1) (\mathcal{A}+5)}{\left(7-\mathcal{A}^2-10 \mathcal{A}\right)^{3/2} (\mathcal{A}+\beta +1)^2}+\frac{6 \mathcal{C}_2}{\sqrt{-\mathcal{A}^2-10 \mathcal{A}+7} (\mathcal{A}+\beta +1)^2}+12 \mathcal{C}_1\Bigg]. \nonumber
\end{eqnarray}

\section{Boundary Conditions}\label{IV}

For a stable object, the stellar distribution must be joined between the exterior and interior spacetimes at the pressure free interface, i.e. where $p_r^{\text{eff}}=0$.  In the present gravity, the Schwarzschild-de Sitter (dS) is more suitable exterior spacetime that can be given by following equation: 
\begin{eqnarray}\label{+}
    dS_{+}^2 &=& -\bigg(1-\frac{2 {\mathcal{M}}}{r}+\frac{\Lambda r^2}{3}\bigg)dt^2+r^2(d\theta^2+\sin^2{\theta} \,d\phi^2)+\bigg(1-\frac{2 {\mathcal{M}}}{r}+\frac{\Lambda r^2}{3}\bigg)^{-1} dr^2.
\end{eqnarray}

The entire mass at the boundary \( r=R \) is denoted by \( M \), whereas \( \Lambda \) signifies the cosmological constant. For smooth joining of the spacetimes (5.1) and (3.14), we must use the first and second fundamental forms. The first fundamental form describes the joining of metric tensor $g_{\alpha\eta}$ components for the spacetime (5.1) and (3.14) at $r=R$, i.e.
\begin{eqnarray}
g_{tt}^{-}|_{r=R}=g_{tt}^{+}|_{r=R}\,, ~~~~~~~\text{and}~~~~
g_{rr}^{-}|_{r=R}=g_{rr}^{+}|_{r=R}\,, \label{eq5.2}
\end{eqnarray}
where $^-$, and $^+$ denote for interior and exterior spacetime. 

The exact formulation is generated using Eq.~(\ref{eq5.2}) as
\begin{eqnarray}\label{C}
    e^{-\mathcal{W}(R)} &=& 1-\frac{2 {\mathcal{M}}}{R}+\frac{\Lambda R^2}{3},\\ \label{c2}
    e^{\mathcal{U}(R)} &=& 1-\frac{2 {\mathcal{M}}}{R}+\frac{\Lambda R^2}{3},
\end{eqnarray}
while the second fundamental form is defined as
\begin{eqnarray}
   p_r^{\text{eff}} (R)=0.
\end{eqnarray}

The measured value of the cosmological constant \(\Lambda \) is \( 10^{-46}/km^2 \), which has a negligible effect on the properties of compact stars. Therefore, we have omitted the cosmological constant \( \Lambda \) from our examination of compact stars. Based on the previously mentioned conditions, we determine the constant parameters as follows:
\begin{eqnarray}
&&\hspace{-0.5cm} \mathcal{C}_2=\Bigg[3 \sqrt{1-\frac{8 \mathcal{B} (\mathcal{B}+3)}{7 (\mathcal{B}+1)^2}} (\sigma +1)^3\Bigg]\Bigg/\Bigg[(\mathcal{B}+\sigma +1)^2 \Bigg(3 \mathcal{C}_3-\Bigg(\frac{\sqrt{7-\mathcal{B} (\mathcal{B}+10)} }{((\sigma -8) \sigma -16) (\mathcal{B}+\sigma +1)^3}\Big(3 (\mathcal{B}\nonumber\\
&&\hspace{0.5cm}-3) \sigma ^4+(\mathcal{B} (\mathcal{B}+16)+119) \sigma ^3+2 (\mathcal{B} (5 \mathcal{B}+86)-47) \sigma ^2+16 (5 (\mathcal{B}-8) \mathcal{B}-53) \sigma -96 (\mathcal{B}+1) \nonumber\\
&&\hspace{0.5cm} \times (3 \mathcal{B}+7)+3 \sigma ^5\Big)+\frac{12 ((\sigma -4) \sigma  (\sigma +8)+128)}{(16-(\sigma -8) \sigma )^{3/2}} \tanh ^{-1}\left(\frac{\mathcal{B} (\sigma -4)+5 \sigma +12}{\sqrt{7-\mathcal{B} (\mathcal{B}+10)} \sqrt{16-(\sigma -8) \sigma }}\right)\Bigg)\nonumber\\
&&\hspace{0.5cm} \times\frac{1}{\left(\sigma ^2-8 \sigma -16\right)^2}\Bigg)\Bigg]^{-1},\\
&&\hspace{-0.5cm} \mathcal{C}_1=\Bigg[3\mathcal{C}_3 \sqrt{1-\frac{8 \mathcal{B} (\mathcal{B}+3)}{7 (\mathcal{B}+1)^2}} (\sigma +1)^3\Bigg]\Bigg/\Bigg[(\mathcal{B}+\sigma +1)^2 \Bigg(3 \mathcal{C}_3-\Bigg(\frac{\sqrt{7-\mathcal{B} (\mathcal{B}+10)} }{((\sigma -8) \sigma -16) (\mathcal{B}+\sigma +1)^3}\Big(3 (\mathcal{B}\nonumber\\
&&\hspace{0.5cm}-3) \sigma ^4+(\mathcal{B} (\mathcal{B}+16)+119) \sigma ^3+2 (\mathcal{B} (5 \mathcal{B}+86)-47) \sigma ^2+16 (5 (\mathcal{B}-8) \mathcal{B}-53) \sigma -96 (\mathcal{B}+1) \nonumber\\
&&\hspace{0.5cm} \times (3 \mathcal{B}+7)+3 \sigma ^5\Big)+\frac{12 ((\sigma -4) \sigma  (\sigma +8)+128)}{(16-(\sigma -8) \sigma )^{3/2}} \tanh ^{-1}\left(\frac{\mathcal{B} (\sigma -4)+5 \sigma +12}{\sqrt{7-\mathcal{B} (\mathcal{B}+10)} \sqrt{16-(\sigma -8) \sigma }}\right)\Bigg)\nonumber\\
&&\hspace{0.5cm} \times\frac{1}{\left(\sigma ^2-8 \sigma -16\right)^2}\Bigg)\Bigg]^{-1}, 
\end{eqnarray}
where the symbol $\mathcal{B}=\mathcal{X} R^2$ (the expression for $ \mathcal{C}_3$ is given in Appendix).

\section{Physical analysis} \label{V}
We have solved the modified gravitational field equation by assuming a physically viable expression for anisotropy as well as that of the radial component of  the metric potential in order to obtain an exact and non-singular solution describing the physical properties of highly dense stellar structures. The variation of the physical properties with respect to the radial distance for varying parameters such as $\psi_1$, $\psi_2$ and $\beta$ are shown graphically. Now, we shall present a physical analysis of these plots in detail in the following sub-cases.

\subsection{Impact of $f(Q,T)$ parameters on density, pressures, and anisotropy}
\begin{figure}[!htp]
    \centering
\includegraphics[height=5.2cm,width=5.3cm]{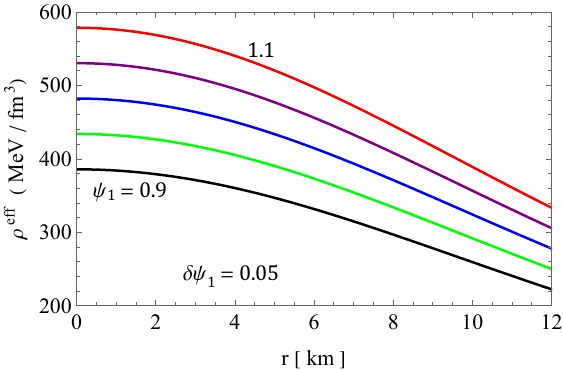}~~\includegraphics[height=5.2cm,width=5.3cm]{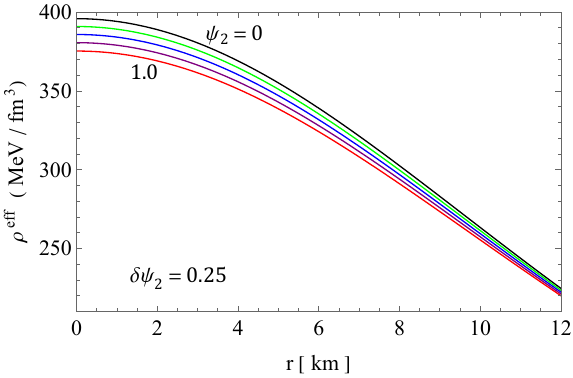}~~\includegraphics[height=5.2cm,width=5.3cm]{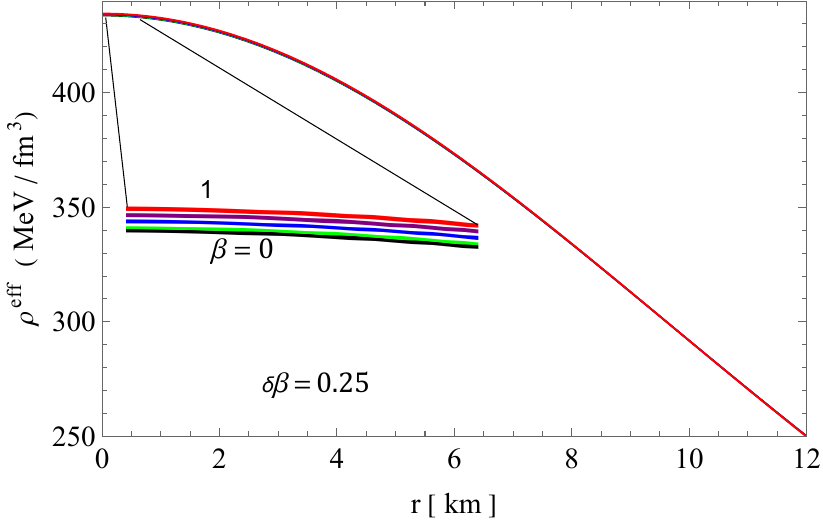}
\caption{Impact of parameters $\psi_1$, $\psi_2$, and $\beta$ on the energy density ($\rho^{\text{eff}}$) against the radial distance $r$ for $\mathcal{X} =0.0016\, \text{km}^{-2}$ with fixed values: (i) $\psi_2=0.5$ and $\beta=0.5$ for left panel, (ii) $\psi_1=0.8$ and $\beta=0.5$ for middle panel, and  (iii) $\psi_1= 0.9$ and $\psi_2=0.5 $ for right panel. }
    \label{f1}  
\end{figure}

Effective density of the present stellar system shows physical behavior from center to the surface for values of varying parameters $\psi_1$, $\psi_2$ and $\beta$ as shown in three panels respectively in Fig. \ref{f1}. It has non-zero finite values at the center and decreasing nature towards the surface for different values of the model parameters. If one increases the value of $\psi_1$, the central density as well as the effective density at each point in the star rise with a significant amount. Since $\psi_1$ parameter is associated to the  non-metricity scalar, it can be inferred that non-metricity has vital role to play in the high dense stars under the background of $f(Q,T)$ gravity. The similar effect of the non-metricity parameter on the energy density has been explored in the studies of compact star in $f(Q)$ gravity \cite{Maurya:2024wtj,Maurya:2024dhk}. Now, increasing values of $\psi_2$ lead to decreasing values of effective density at every point inside the star. Analogous impact of the parameter associated to $T$ on the effective density can be seen in $f(R,T)$ gravity \cite{Maurya:2025kto}. Additionally, the rate of change of effective density with respect to $\psi_1$ at a particular radial coordinate is larger than the that with respect to $\psi_2$ at the same point. So, left panel and middle panel of  Fig. \ref{f1} signify that the effect of $\psi_1$ on effective density runs counter to the impact of parameter $\psi_2$ on the changes in the effective density. This kind of feature has been explored in a recent work \cite{Maurya:2025psm} on charged compact star under $f(Q,T)$ gravity. The right panel of Fig. \ref{f1} suggests that increasing values of $\beta$ have an insubstantial effect on the positive change in the effective density everywhere inside the star.

\begin{figure}[!htp]
    \centering
\includegraphics[height=5.2cm,width=5.3cm]{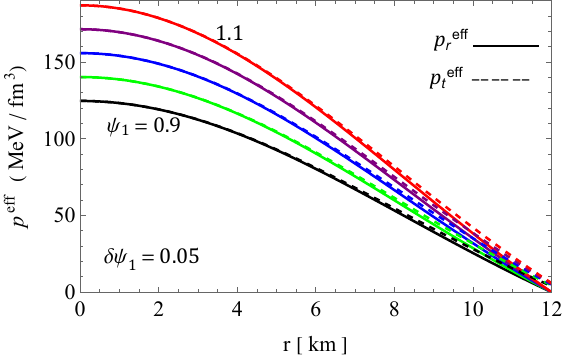}~~\includegraphics[height=5.2cm,width=5.3cm]{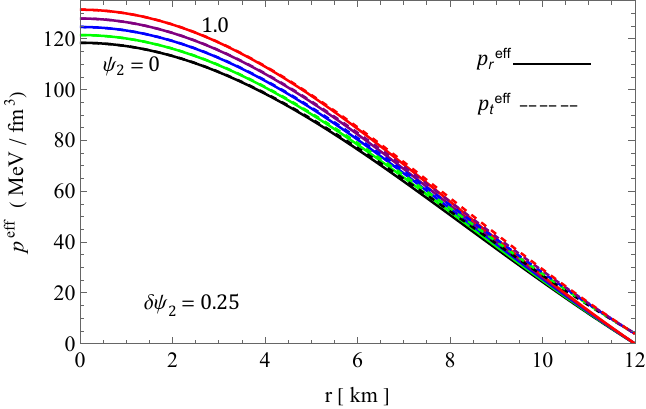}~~\includegraphics[height=5.2cm,width=5.3cm]{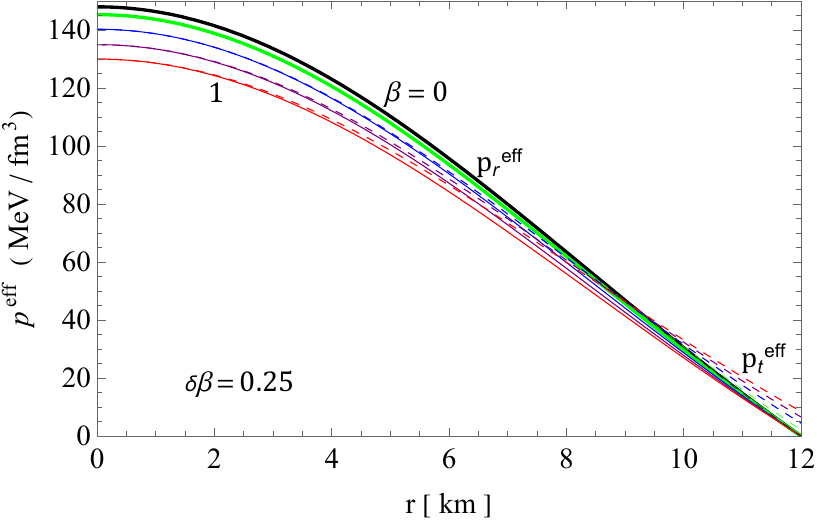}
    \caption{Impact of parameters $\psi_1$, $\psi_2$, and $\beta$ on the radial and tangential pressures ($p^{\text{eff}}_r$ \& $p^{\text{eff}}_t$) against the radial distance $r$. For these plots, we select the same data set of values as used in Fig.~\ref{f1}.}
    \label{f2}
\end{figure}

The tangential pressure and radial pressure show identical variations throughout the star. This can be seen from the graphical presentations as shown in Fig. \ref{f2}. Interestingly, both the pressures have the exactly same value at the center of the star leading to the zero central anisotropy which is considered to be one of the accepted physical requirements for a physically valid anisotropic star. Another physical criterion known as {\it vanishing of the radial pressure at the surface of the star} is fulfilled for different values of the model parameters. It can be noted that the tangential pressure are slightly larger than the radial pressure near the surface as it can be seen from the panels of Fig. \ref{f2}. 

Both the pressures increase more in magnitude around the central region than that in magnitude near the surface for increasing values of parameters $\psi_1$ and $\psi_2$. However, the pressures show decreasing trend at the central region for increasing $\beta$. As $\beta$ approaches to zero the pressures in the radial and the tangential direction  tend to become equal in magnitude and identical in nature which typically represents the isotropic nature of the stellar system. This is in agreement with the fact that the anisotropy becomes zero for the vanishing $\beta$ as indicated in the expression of $\Delta^{\text{eff}}$. 

\begin{figure}[!htb]
    \centering
\includegraphics[height=5.2cm,width=5.3cm]{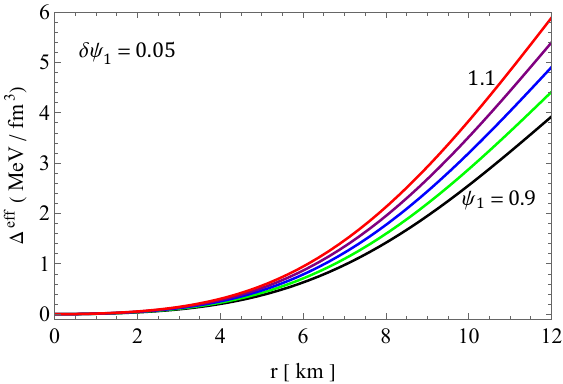}~~\includegraphics[height=5.2cm,width=5.3cm]{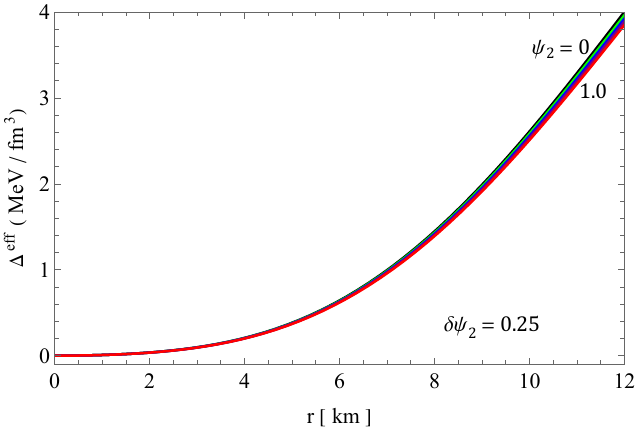}~~\includegraphics[height=5.2cm,width=5.3cm]{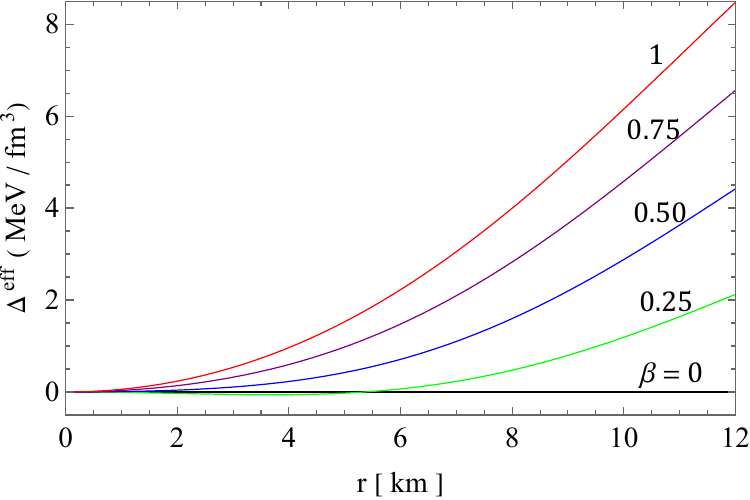}
    \caption{Impact of parameters $\psi_1$, $\psi_2$, and $\beta$ on the anisotropy ($\Delta^{\text{eff}}$) against the radial distance $r$. For these plots, we select the same data set of values as used in Fig.~\ref{f1}. }
    \label{f3}
\end{figure}


The variation of anisotropy is illustrated graphically in the panels of Fig. \ref{f3}. It is evident that the zero central anisotropy is achieved in the present stellar model under $f(Q,T)$ gravity which is in accordance with the results obtained graphically in Fig. \ref{f2}. Further, the anisotropy shows increasing nature with regard to the radial distance for various values of parameters while its value becomes maximum at the surface. The impact of parameter $\psi_1$ on the anisotropy is more prominent as compared to the effect of parameter $\psi_2$ on the anisotropy. Now, $\beta$ being the anisotropy parameter, it eventually tunes the anisotropy within the stellar structure as shown in the right panel of Fig. \ref{f3}.

\subsection{Causality Condition} 
Now, we examine whether the present stellar system in the framework of $f(Q,T)$ gravity maintains the causality condition which indicates that sound speeds should be positive and less than the speed of light. This is significant in connection to the stability analysis of the present anisotropic stellar configuration. The sound speeds in radial direction ($v_r$) as well as the tangential direction ($v_t$) can defined as 
  \begin{equation}
      v_r^2 =  \frac{dp_r^{\text{eff}}}{d\rho^{\text{eff}}} ,\quad \quad v_t^2=  \frac{dp_t^{\text{eff}}}{d\rho^{\text{eff}}}.
  \end{equation}

\begin{figure}[!htp]
    \centering
\includegraphics[height=5.3cm,width=5.3cm]{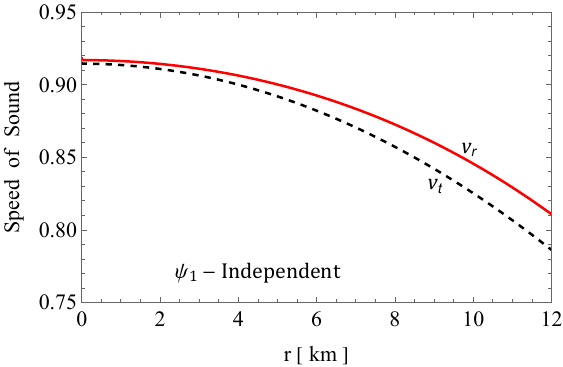}~~\includegraphics[height=5.2cm,width=5.3cm]{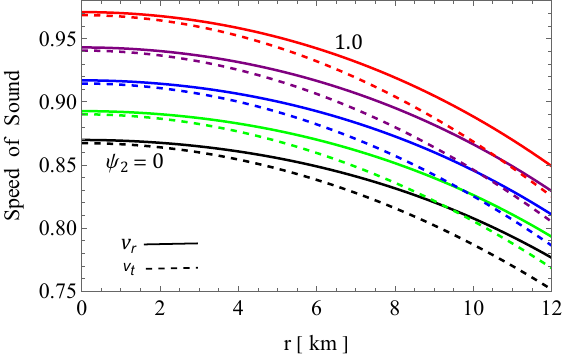}~~\includegraphics[height=5.2cm,width=5.3cm]{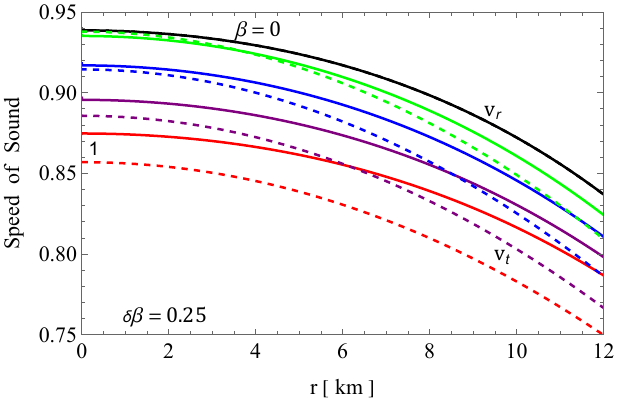}
    \caption{Impact of parameters $\psi_1$, $\psi_2$, and $\beta$ on the velocity of sounds ($\text{v}^2_r$ \& $\text{v}^2_t$) against radial distance $r$ for $\mathcal{X} =0.0016\, \text{km}^{-2}$ with fixed values: (i) $\psi_2=0.5$ and $\beta=0.5$ for left panel, (ii) $\psi_1=0.8$ and $\beta=0.5$ for middle  panel, and  (iii) $\psi_1= 0.9$ and $\psi_2=0.5 $ for right  panel. }
    \label{f4}
\end{figure}

The nature of change in the sound speeds with respect to the radial distance and varying parameters is shown in the panels of Fig. \ref{f4}. The analysis of these graphical plots confirms that causality condition, i,e,, $0\leq v^2_r \leq 1$ and  $0\leq v^2_t \leq 1$ is satisfied everywhere inside the star under the variation of the $f(Q,T)$ parameters and anisotropy parameter. 

\subsection{Stability analysis via adiabatic index}

Let us adopt another approach to analyze the stability of the  present anisotropic stellar system in terms of examining the lower bounds on the  adiabatic index ($\Gamma$) expressed as 
\begin{equation}
    \Gamma = \frac{\rho^{\text{eff}}+P_r^{\text{eff}}}{P_r^{\text{eff}}}\frac{dP_r^{\text{eff}}}{d\rho^{\text{eff}}}. 
\end{equation}

An anisotropic stellar system will be stable under the necessary condition that the adiabatic index should obey the following inequality given as 
\begin{equation}
\Gamma > \frac{4}{3}\left(1 + \frac{\Delta}{r|(P_{r}^{\text{eff}})|^{\prime}}+\frac{1}{4}\frac{\kappa \rho^{\text{eff}} P_{r}^{\text{eff}}r}{|(P_{r}^{\text{eff}})|^{\prime}} \right).\label{Gamma}
\end{equation}

Note that this is  departed from the stability criterion~\cite{chan1992dynamical,chan1993dynamical} on the adiabatic index given by Chandrasekhar \cite{chandrasekhar1964dynamical} in the case of isotropic  stellar system. It is to be mentioned that  the Newtonian limit  for an isotropic system is given by $\Gamma > \frac{4}{3}$, \cite{heintzmann1975neutron,bondi1992anisotropic}. In Eq. (\ref{Gamma}), the symbol prime denotes  derivative of a function with regard to the radial distance whereas the second and third terms are used to include adjustments corresponding to the anisotropy and relativistic correction.

\begin{figure}[!htp]
    \centering
\includegraphics[height=5.2cm,width=5.3cm]{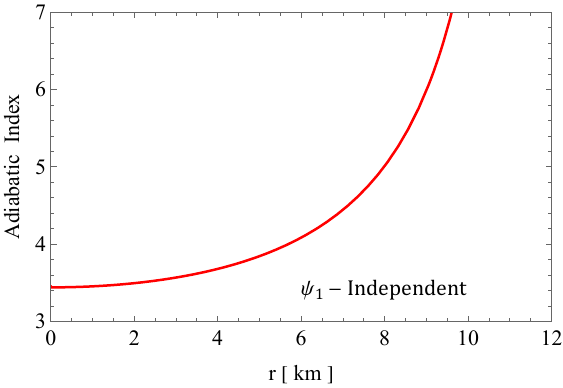}~~\includegraphics[height=5.2cm,width=5.3cm]{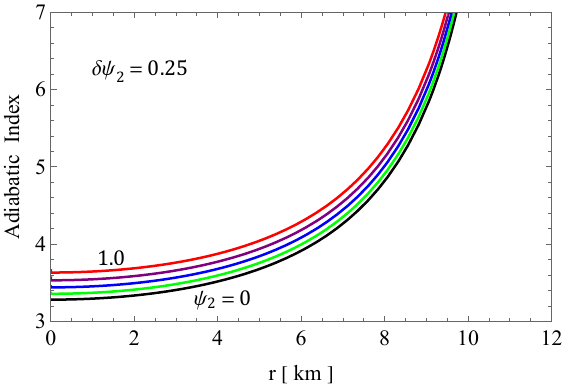}~~\includegraphics[height=5.2cm,width=5.3cm]{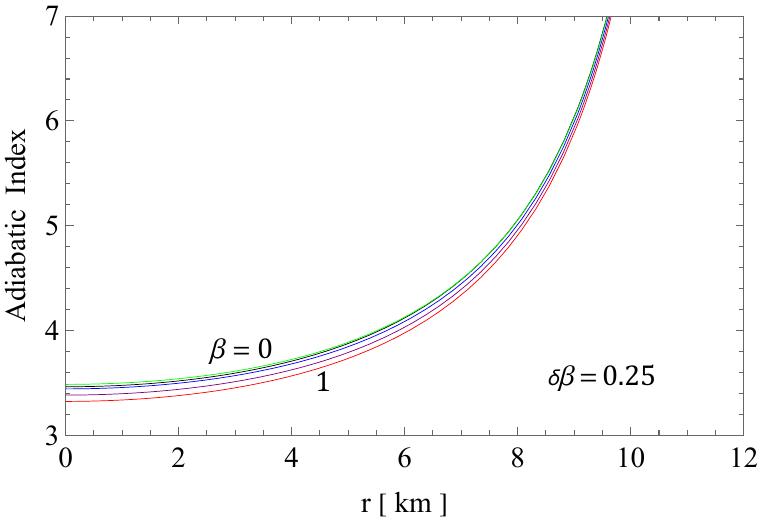}
    \caption{Impact of parameters $\psi_1$, $\psi_2$, and $\beta$ on the adiabatic index ($\Gamma$) against radial distance $r$ $\mathcal{X} =0.0016\, \text{km}^{-2}$ with fixed values: (i) $\psi_2=0.5$ and $\beta=0.5$ for left panel, (ii) $\psi_1=0.8$ and $\beta=0.5$ for middle panel, and  (iii) $\psi_1= 0.9$ and $\psi_2=0.5 $ for right panel. }
    \label{f5}
\end{figure}

The graphical nature of $\Gamma$ with regard to radial distance have been shown in three panels of Fig. \ref{f5} corresponding to the variations in model parameters. The adiabatic index has increasing trend within the anisotropic stellar system as shown in each panel. The left panel interprets that the adiabatic index is independent of the parameter $\psi_1$. It signifies that the non-metricity has no direct influence on the adiabatic condition of stability. Since the pressures become equal in magnitude at the center, the central value of $\Gamma$ should exceed   4/3 for the stable system. In the present anisotropic system, the central $\Gamma$ is greater than three and it increases with high slope. This ensures that the present stellar system is stable under the adiabatic condition. The increasing values of $\psi_2$ enhance the value of $\Gamma$ everywhere inside the star as indicated from the middle panel in Fig. \ref{f5}. On the other hand, increasing values of anisotropy parameter decrease the values of $\Gamma$ for each radial distance as depicted in the right panel in Fig. \ref{f5}. The impact of parameters $\psi_2$ and $\beta$ on the $\Gamma$ is smaller at the surface in comparison with that at the central region of star.

\subsection{Hydrostatic balance via modified Tolman-Oppenheimer-Volkoff equation}
It is necessary to investigate the Tolman-Oppenheimer-Volkoff (TOV) equation~\cite{Tolman1939,OV1939} for $f(Q,T)$ gravity in order to analyze the hydrostatic equilibrium balance of the present stellar configuration under $f(Q,T)$ gravity.  The modified TOV equation in $f(Q,T)$ gravity can be expressed  by the following equation:  
 \begin{eqnarray}
   &&\hspace{-1cm}   - \frac{\mathcal{U}^\prime}{2} (\rho^{\text{eff}}+p_r^{\text{eff}})-\frac{dp_r^{\text{eff}}}{dr}+\frac{2}{r}(p_t^{\text{eff}}-p_r^{\text{eff}})-\frac{\psi_2}{6(8\pi+\psi_2)} \Big[3\frac{d\rho^{\text{eff}}}{dr}-5 \frac{dp_r^{\text{eff}}}{dr}-10 \frac{dp_t^{\text{eff}}}{dr}\Big]=0. \label{eq6.1}
\end{eqnarray}

\begin{figure}[!htp]
\centering
\includegraphics[height=5.2cm,width=5.3cm]{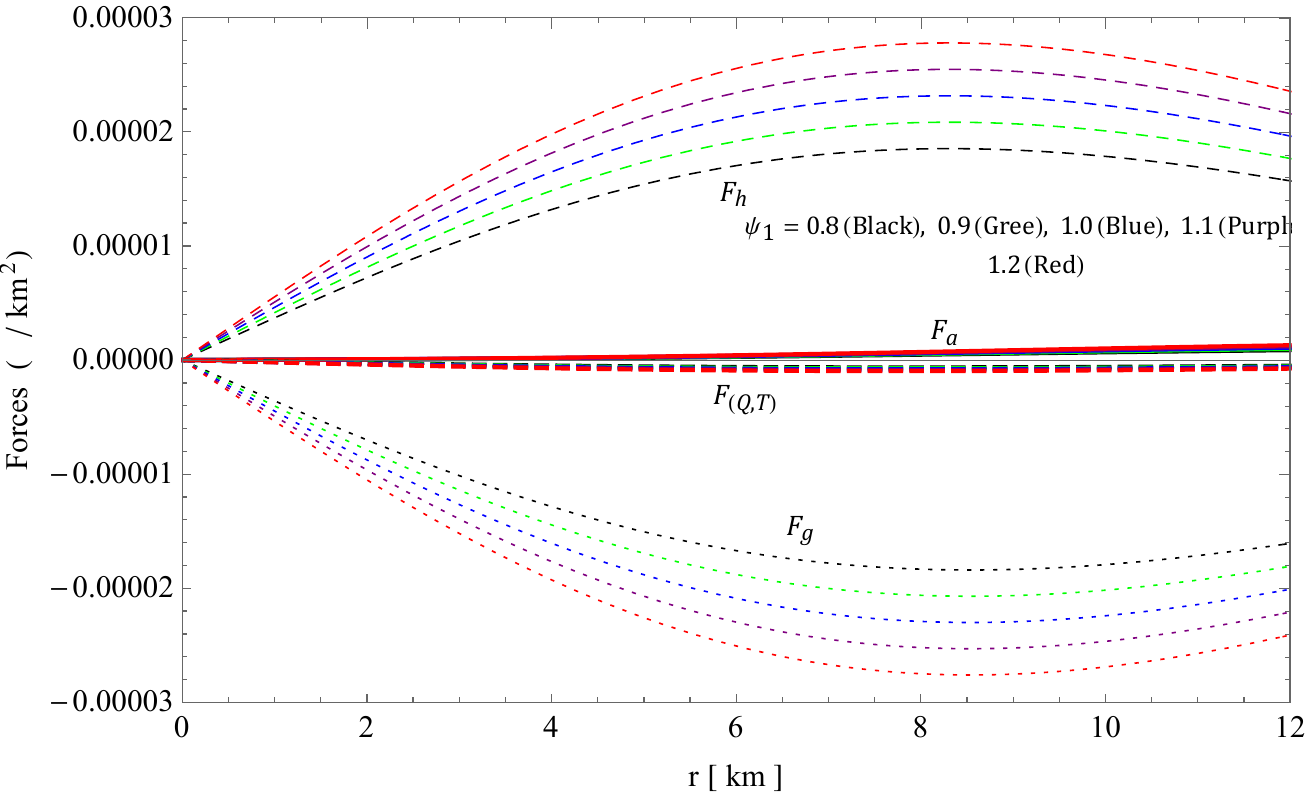}~~\includegraphics[height=5.2cm,width=5.3cm]{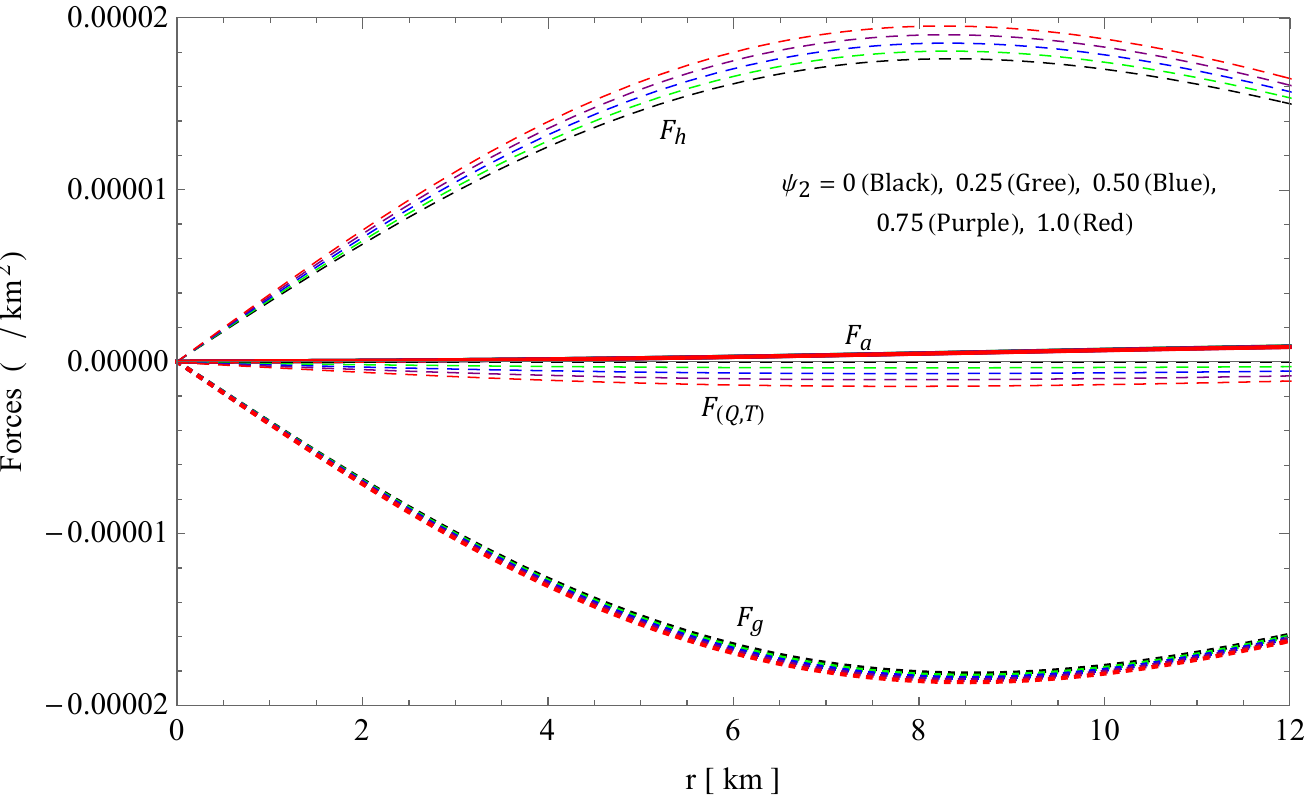}~~\includegraphics[height=5.2cm,width=5.3cm]{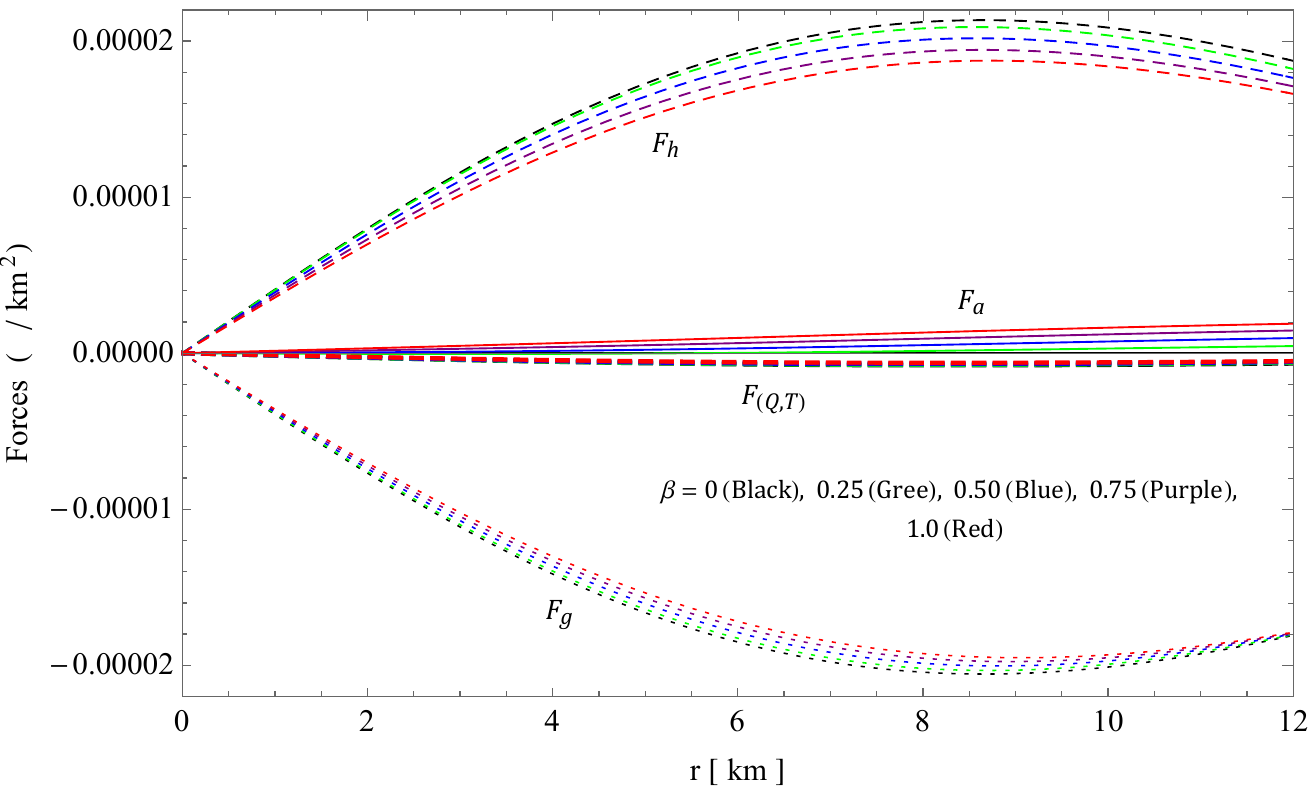}
\caption{Impact of parameters $\psi_1$, $\psi_2$, and $\beta$ on the forces acted on the system ($\Gamma$) against radial distance $r$ $\mathcal{X} =0.0016\, \text{km}^{-2}$ with fixed values: (i) $\psi_2=0.5$ and $\beta=0.5$ for left panel, (ii) $\psi_1=0.8$ and $\beta=0.5$ for middle panel, and  (iii) $\psi_1= 0.9$ and $\psi_2=0.5 $ for right panel. }
   \label{f5tov} 
\end{figure}

The above Eq. (\ref{eq6.1}) can be represented in terms of forces with physical implications given as
\begin{eqnarray}
F_g=- \frac{\mathcal{U}^\prime}{2} (\rho^{\text{eff}}+p_r^{\text{eff}}), ~~F_h=-\frac{dp_r^{\text{eff}}}{dr},~~ F_a=\frac{2 (p_t^{\text{eff}}-p_r^{\text{eff}})}{r},~~\nonumber\\F_{(Q,T)}=-\frac{\psi_2}{6(8\pi+\psi_2)} \Big[3\frac{d\rho^{\text{eff}}}{dr}-5 \frac{dp_r^{\text{eff}}}{dr}-10 \frac{dp_t^{\text{eff}}}{dr}\Big],
\end{eqnarray}
where $F_g$, $F_h$, $F_a$ and $F_{(Q,T)}$ are the forces representing gravitational, hydrostatic, anisotropic and non-minimal coupled type interactions respectively. We have plotted these forces with respect to the radial distance in the panels of Fig. \ref{f5tov} for varying parameters to examine the nature of these interacting forces. Upon evaluation of the panels of Fig. \ref{f5tov} we find that hydrostatic and anisotropic forces are positive and repulsive in nature whereas gravitational and coupling forces are negative and attractive in nature. Interestingly, the contrary nature of these forces helps the stellar system to be stabilized with equilibrium by neutralizing the net combined effect of these forces.

\subsection{Static stability analysis via Harrison-Zel'dovich-Novikov criterion}
Chandrasekhar~\cite{chandrasekhar1964dynamical} had shown, in one of his earlier works, a method to obtain the stable equilibrium of a stellar system under the radial perturbations. In this circumstance, the physical quantities such as the effective density and the pressures along with the metric potentials got perturbed showing oscillatory nature with  the characteristic frequency. This can be expressed mathematically by a  characteristic equation which indicates that  the positivity of the  characteristic frequency can prevent gravitational collapse under a small radial perturbation. Considering the polytropic EOS \cite{harrison1965gravitation} and \cite{zel1972relativistic} with $\Gamma>4/3$ and the positivity of the characteristic frequency  Chandrasekhar proved that the total mass of a stellar should be increasing function of the central density to attain stable equilibrium under a finite radial perturbation. Eventually, the slope of $M$ vs $\rho_c$ curve should be positive,  $dM/d\rho_c >0$, which is recognized as the static stability criteria (or also called as the Harrison-Zel'dovich-Novikov criterion).   

So, we have represented the mathematical expressions of the physical quantities $M,~ \rho_c,$ $dM/d\rho_c$ in the following equations to investigate stability of stellar system as
\begin{eqnarray}
M(\rho_c) &=& \frac{4 \mathcal{X} R^3 (\mathcal{X}R^2+3)}{7 \left(1+\mathcal{X}R^2\right)^2},
\end{eqnarray}
where
\begin{eqnarray}
\mathcal{X} &=& \frac{56 \pi  \left(4 \pi -\psi _2\right) \left(\psi _2+8 \pi \right) \rho _c}{\psi _1} ~\Bigg[288\pi +\psi_2 \Bigg\{-24-\frac{70}{\beta +1}-\frac{15 \sqrt{7}\,\mathcal{C}_2}{(\beta +1)^4} \nonumber \\
&& \Bigg(3 \mathcal{C}_1+\frac{\sqrt{7} \left(-3 \beta ^5+9 \beta ^4-119 \beta ^3+94 \beta ^2+848 \beta +672\right) \mathcal{C}_2}{(\beta +1)^3 \left(\beta ^2-8 \beta -16\right)^3}-\nonumber \\
&& \frac{12 \left(\beta ^3+4 \beta ^2-32 \beta +128\right) \mathcal{C}_2 }{\left(16-\beta ^2+8 \beta\right)^{7/2}}~\tanh ^{-1}\left(\frac{5 \beta +12}{\sqrt{112-7 (\beta -8) \beta }}\right)\Bigg)^{-1}\Bigg\} \Bigg]^{-1}
\end{eqnarray}
   
\begin{figure}
\centering
\includegraphics[height=5.2cm,width=5.3cm]{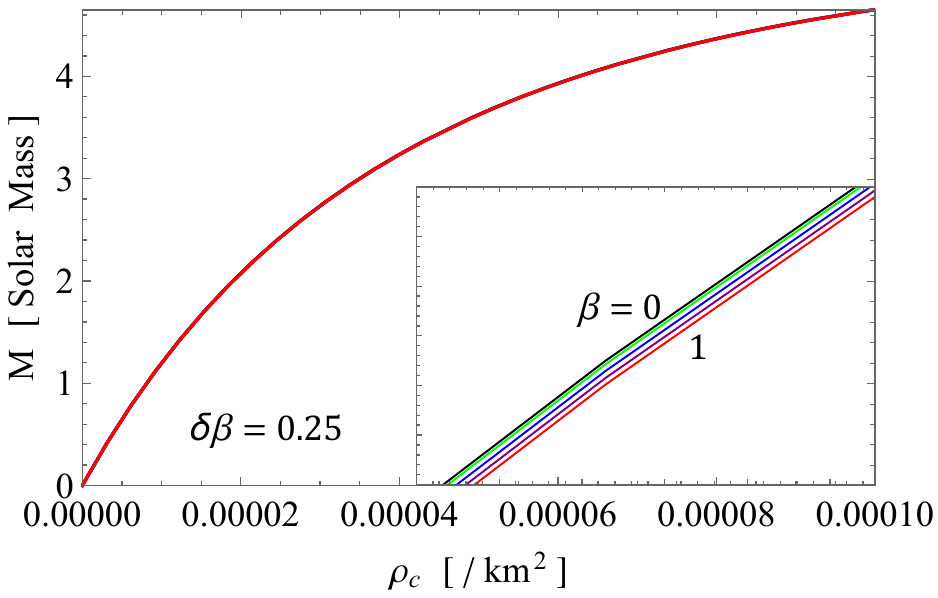}~~\includegraphics[height=5.2cm,width=5.3cm]{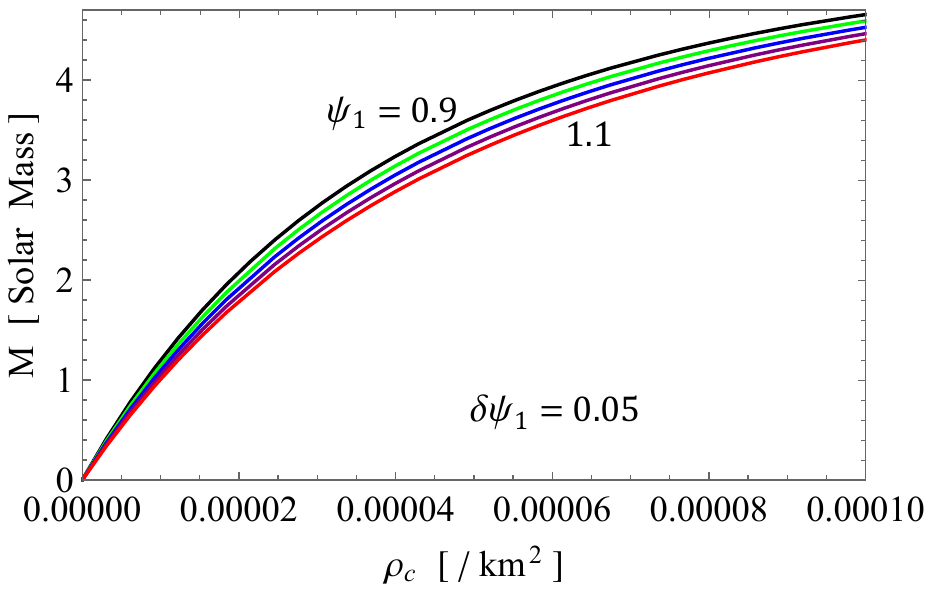}~~\includegraphics[height=5.2cm,width=5.3cm]{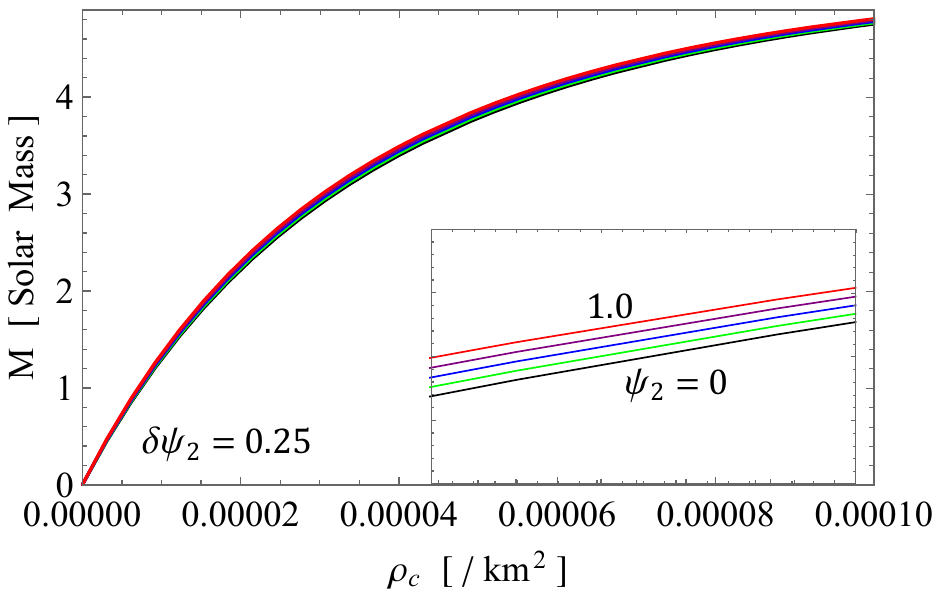}
\caption{Impact of parameters $\psi_1$, $\psi_2$, and $\beta$ on the $M-\rho_c$ for fixed values: (i) $\psi_1= 0.9$ and $\psi_2=0.5 $ for left panel, (ii) $\psi_2=0.5$ and $\beta=0.5$ for middle panel, and  (iii) $\psi_1=0.8$ and $\beta=0.5$ for right panel. }
    \label{f5mro}
\end{figure}

The variation of the total mass with respect to the central density have been presented for the varying parameters in the three panels of Fig. \ref{f5mro}. Apparently, each panel suggests that the total mass has a monotonically increasing nature with regard to the central density. The impact of the parameters on the $M-\rho_c$ curves is seen to be nominal in magnitude. However, a thorough inspection of the panels of Fig. \ref{f5mro} probes that the increasing values of $\beta$ and $\psi_1$  but the decreasing values of $\psi_2$ reduce the slope of the $M-\rho_c$ curves to a small extent.

\subsection{Astrophysical consequences of $f(Q,T)$ gravity on mass–radius relations}
We have considered some observed massive pulsars of mass exceeding 2  $M_{\odot}$  in order to investigate the impact of anisotropy and the model parameters on the properties of highly dense stars and mass-radius relation in the context of $f(Q,T)$ gravity.  Now, the neutron stars transmitting regular and high intensity electromagnetic bursts are essentially known as the pulsars.  The bursts are generated mainly due to the presence of intense magnetic fields in the rotating pulsars. The high compactness as well as high density of such pulsars are indicated by the rapid rotations with high angular frequencies. Certainly, the stability of such type pulsars can be sustained by the EOSs which are stiffer in nature. Again, the relation between mass and radius of a compact star provides some evidences to unveil the unknown properties of highly dense matter which is governed by the stiffer EOS. Therefore, it become necessary to examine the mass-radius relation in the present model under  $f(Q,T)$ gravity.  

In the present study  PSR J074 + 6620 \cite{r1-PSRJ074+6620} and PSR J1810+1744 \cite{r1-PSRJ1810+1744-1} which are  binary systems consist of high mass pulsar with  a companion star of comparatively low mass have been taken into consideration. Additionally, another stellar system having the black window, i.e.,  PSR J1959+2048 \cite{star1} and  the redback, i.e., PSR J2215+5135  \citep{star1} are also taken into account. Moreover, we have included  GW190814 \cite{waves3} in the current analysis of the mass-radius relation in connection to the recent developments in GW observations. Nevertheless, the estimation of the values of radii of neutron stars is more intricate \cite{Fortin2015,Ozel2016a} to obtain with regard to observational parameters in comparison with  the measurements of masses of the observed stars. In this context, the radiation radius defined as $R_{\infty} = R \left(1- \frac{2M}{R}\right)^{-1/2}$ considered to be reliable means to estimate the radius of neutron star which is less than $R_{\infty}$.

Recently, measurements of the size of hotpots \cite{Miller2019,Riley2021} on the surfaces of neutron star is pioneered by NICER  which shows a new way of determining radius of a neutron star.  In a study \cite{Baubock:2015ixa}, size of hotspot is estimated for sources of various categories and the physical implications arising from the consideration of small size of spot have been discussed . In general, a local hotspot is recognized as a small and localized region on the surface of a neutron-star typically located around its magnetic pole. This is generated due to the heat produced by returning magnetic currents and the temperature anisotropy observable in the pulse profile of a star. By investigating the patterns of rotating surface of neutron star, the estimated radius of PSR J0740+6620 is found as $13.70^{+2.6}_{-1.5}$ km  \citep{Miller:2021qha}  with  68\% credibility. Again, the radius of PSR J0740+6620 has been estimated more precisely in an another study \cite{Dittmann:2024mbo}  to be  $ 12.92_{-1.13}^{+2.09}$ km, i.e., lying within the range \{11.79 km, 15.01 km\}  by investigating NICER and XMM-Newton X-ray observational data.

\begin{figure}
    \centering
\includegraphics[height=10cm,width=12.5cm]{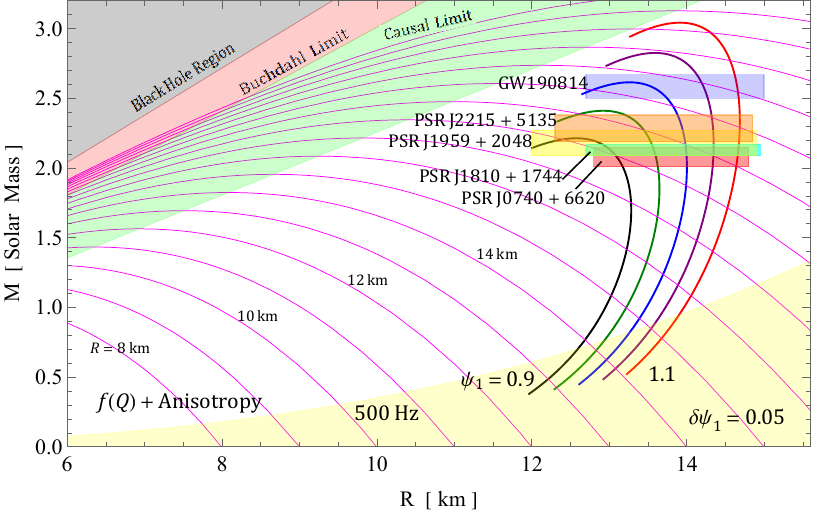}\\ \includegraphics[height=10cm,width=12.5cm]{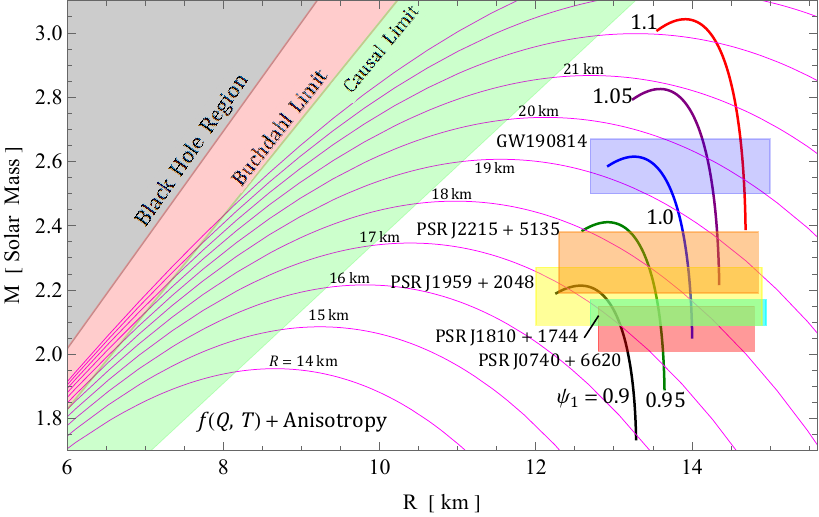}
\caption{$M-R$ curves for $\mathcal{X} = 0.0016 \,\text{km}^{-2}$ with fixed values: (i) $\psi_2=0$ and $\sigma=0.01$ for top panel and (ii) $\psi_2=0.01$ and $\sigma=0.01$ for bottom panel. }
    \label{f6} 
\end{figure}

\begin{figure}
    \centering
\includegraphics[height=10cm,width=12.5cm]{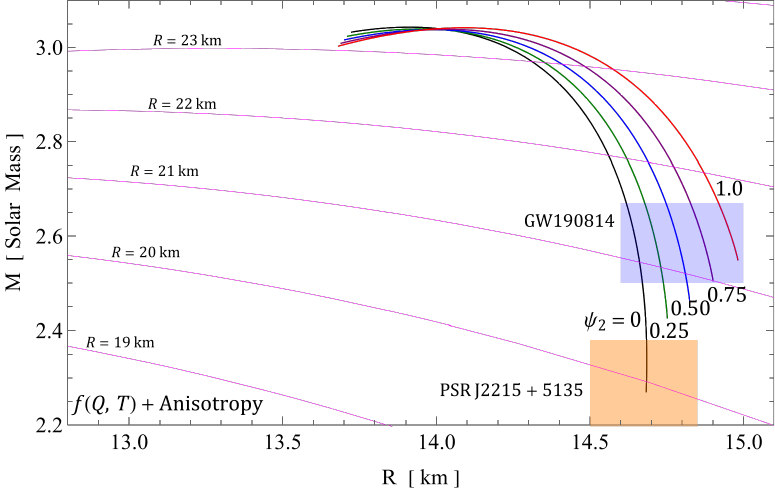}\\\includegraphics[height=10cm,width=12.5cm]{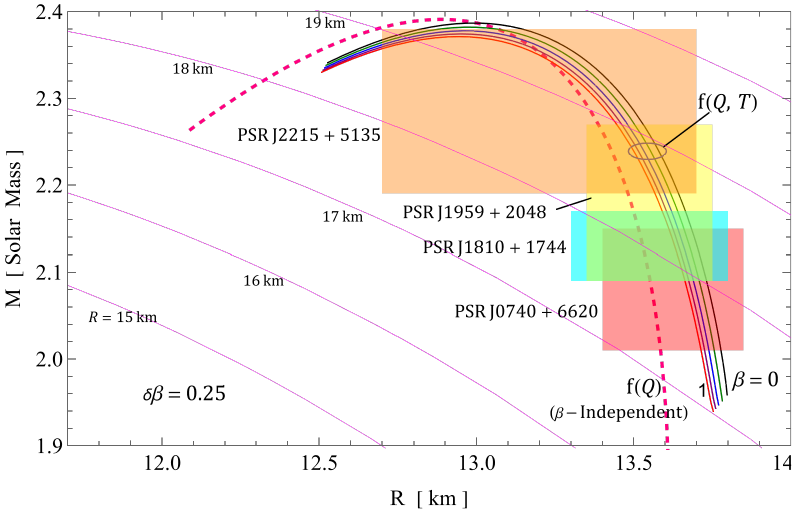}
\caption{$M-R$ curves for $\mathcal{X} = 0.0016 \,\text{km}^{-2}$ with fixed values: (i) $\psi_1=1.1$ and $\sigma=0.01$ for top panel and (ii) $\psi_2=1.05,\,\psi_2=0.7$ [$f(Q,T)$] and $\psi_2=1.05,\,\psi_2=0$ [$f(Q)$] for bottom panel.}
    \label{f7} 
\end{figure}

In the present investigation we have constructed gravitational models to study mass-radius relation for pulsars assuming a well-behaved metric potential and pressure anisotropy function in the framework of $f(Q,T)$ gravity. The mass-radius curves are represented in Fig.~\ref{f6} for various values of $\psi_1$ with constant value of anisotropy parameter where the top panel is for $\psi_2=0$ implying the $f(Q)$ gravity case and the bottom panel is for $\psi_1\neq0$) indicating $f(Q,T)$ gravity case.  Mass-radius curves are plotted in the two panels of Fig.~\ref{f6} to examine any change of the impact of $\psi_1$ parameter on the $MR$ curves due to non minimal coupling of $Q$ and $T$. By comparing the two panels we see that the impact of $\psi_1$ on $MR$ curves is similar for both cases for $\psi_1\leq1.0$. On the other hand, $M-R$ curves for $\psi_1>1.0$ can not predict the radius for less massive stars because of the increase in nuclear density beyond the maximum limit in case of $f(Q,T)$ gravity. This type of trend in $M-R$ curves can be seen in a recent work \cite{Saha:2024swd} which investigated mass-radius relationships for all possible types of EOSs with $M_{max}$ greater than 2 $M_{\odot}$ for the non rotating stellar model  and explored that the less massive stars tends to have its maximum masses attaining cores with higher densities relative to the cores of comparatively highly massive stars. In both the panels of Fig. \ref{f6}, the $M-R$ curves the peak or $M_{max}$ and the corresponding radius increase due to increase in $\psi_1$.  The mass-radius relation with regard to varying $\psi_2$ for fixed values of $\psi_1$ and $\beta$ has been represented in the top panel of Fig. \ref{f7}. The $M_{max}$ from the each $MR$ curve remains almost same for increasing $\psi_2$ whereas the radius corresponding to the $M_{max}$ increases as shown in the Fig. \ref{f7}. Further, the $M-R$ curves tend to converge to a certain point as can be observed in  Fig. \ref{f7} depicting the fixed value of the mass and the radius of that intersecting point independent on the impact of $\psi_2$. 

In the bottom panel of Fig. \ref{f7}, the effect of anisotropy on the mass-radius relation have been shown graphically. Notably, the impact of anisotropy on the physical properties of stellar system depends on the type of gravitational model and the quantity of anisotropy introduced in that model. It has been shown in a study \cite{Silva:2014fca} that the observational data related to binary pulsars could restrict the range of anisotropy. Based on relativistic mean field theory, another study \cite{Rahmansyah:2021gzt} found consistency of multi-messenger observational data with the imposed restrictions on the neutron star. Further, observational data related to the tidal deformability arising from the merger event GW170817 can constrain the anisotropy range \cite{Biswas:2019gkw}. Likewise, observations from merger events like GW170817 and GW190814 indicate that various amount of anisotropy could constrain the values of $f$-mode frequency as well as the moment of inertia \cite{Das:2022ell,Mohanty:2023hha}. Again, limitations on the range of anisotropy can be put by the NICER observations of estimated mass and radius at equator of PSR J0030+045 \cite{Riley:2019yda} and PSR J0740+662 \cite{Riley:2021pdl}. 

The values of $M_{max}$ obtained from the peak of $M-R$ curves shown in Fig. \ref{f7} are seen to be increasing for decreased values of $\beta$. This implies that the increased amount of anisotropy decreases $M_{max}$ of the present stellar configuration with nominal effect of $\beta$ on the radius corresponding to $M_{max}$ in the framework of $f(Q,T)$ gravity. Additionally, the $M-R$ curve  which is independent of the impact of $\beta$ in the framework of $f(Q)$ gravity has been shown in the bottom panel of Fig. \ref{f7}. This particular curve of $M-R$ in  $f(Q)$ gravity when compared with the $M-R$ curve with $\beta=0$ in $f(Q,T)$ gravity, we note that the $M_{max}$ has been raised while the radius is decreased in case of $f(Q)$ gravity. This confirms that the stellar structure with the higher compactness and density for zero anisotropy is more preferable in the background of $f(Q)$ gravity relative to that in  $f(Q,T)$ gravity.

\begin{table*}[!htp]
\centering
\caption{The predicted radii of a few high mass compact stars  in  $f(Q)$  gravity (from Fig. \ref{f6}).}\label{tab1}
\begin{minipage}{0.8\textheight}
 \scalebox{0.9}{\begin{tabular}{| *{7}{c|} }
\hline
 &    & \multicolumn{5}{c|}{Predicted $R$ (km) ~~$f(Q)$}  \\[0.15cm]
\cline{3-7}
{Objects} & {$\frac{M}{M_\odot}$} & \multicolumn{5}{c|}{ $\psi_1$}  \\[0.15cm]
\cline{3-7}
&  & $0.90$ & $0.95$ & $1.00$ & $1.05$ & $1.10$  \\[0.15cm] \hline
PSR J074+6620 \citep{r1-PSRJ1810+1744-1} & 2.08$\pm$0.07  & $13.09_{-0.12}^{+0.08}$  &  $13.60_{-0.04}^{+0.03}$  &   $14.00_{-0.01}^{+0.01}$  &  $14.34_{-0.01}^{+0.01}$  &  $14.66_{-0.01}^{+0.02}$     \\[0.15cm]
\hline
PSR J1810+1744 \citep{r1-PSRJ1810+1744-1} & 2.13$\pm$0.04  & $13.01_{-0.10}^{+0.07}$  &  $13.57_{-0.02}^{+0.03}$  &   $13.99_{-0.01}^{+0.01}$  &  $14.35_{-0.01}^{+0.01}$  &  $14.67_{-0.01}^{+0.01}$     \\[0.15cm]
\hline
PSR J1959+2048 \citep{star1} & 2.18$\pm$0.09  & $12.88_{-}^{+0.10}$  &  $13.54_{-0.09}^{+0.06}$  &   $13.98_{-0.03}^{+0.02}$  &  $14.35_{-0.01}^{+0.01}$  &  $14.67_{-0.01}^{+0.01}$    \\[0.15cm]
\hline
PSR J2215+5135 \citep{star1} & $2.28^{+0.10}_{-0.09}$ &  $-$  &  $13.43_{-0.22}^{+0.10}$  &   $13.94_{-0.06}^{+0.04}$  &  $14.34_{-0.02}^{+0.01}$  &  $14.68_{-0.01}^{+0.01}$   \\[0.15cm]
\hline
GW190814 \citep{waves3} & 2.5-2.67 &  $-$  &   $-$  &  $13.58_{-}^{+0.27}$  &  $14.23_{-0.10}^{+0.04}$  &   $14.65_{-0.03}^{+0.02}$    \\[0.15cm]
\hline
\end{tabular}}
 \end{minipage}
\end{table*}
\begin{table*}[!htp]
\centering
\caption{The predicted radii of a few high mass compact stars  in  $f(Q,T)$ gravity (from Fig. \ref{f6}).}\label{tab2}
\begin{minipage}{0.8\textheight}
 \scalebox{0.9}{\begin{tabular}{| *{7}{c|} }
\hline
 &     & \multicolumn{5}{c|}{Predicted $R$ (km) ~~$f(Q,T)$}  \\[0.15cm]
\cline{3-7}
{Objects} & {$\frac{M}{M_\odot}$}  & \multicolumn{5}{c|}{$\psi_1$} \\[0.15cm]
\cline{3-7}
&   & $0.90$ & $0.95$ & $1.00$ & $1.05$ & $1.10$ \\[0.15cm] \hline
PSR J074+6620 \citep{r1-PSRJ1810+1744-1} & 2.08$\pm$0.07   &   $13.09_{-0.12}^{+0.08}$  &  $13.60_{-0.04}^{+0.03}$  &   $14.00_{-0.01}^{+0.01}$  &  $-$  &  $-$   \\[0.15cm]
\hline
PSR J1810+1744 \citep{r1-PSRJ1810+1744-1} & 2.13$\pm$0.04   &   $13.01_{-0.10}^{+0.07}$  &  $13.57_{-0.02}^{+0.03}$  &   $13.99_{-0.01}^{+0.01}$  &  $-$  &  $-$   \\[0.15cm]
\hline
PSR J1959+2048 \citep{star1} & 2.18$\pm$0.09  &   $12.88_{-}^{+0.10}$  &  $13.54_{-0.09}^{+0.06}$  &   $13.98_{-0.03}^{+0.02}$  &  $14.35_{-0.01}^{+0.01}$  &  $-$  \\[0.15cm]
\hline
PSR J2215+5135 \citep{star1} & $2.28^{+0.10}_{-0.09}$  &   $-$  &  $13.43_{-0.22}^{+0.10}$  &   $13.94_{-0.06}^{+0.04}$  &  $14.34_{-0.02}^{+0.01}$  &  $-$  \\[0.15cm]
\hline
GW190814 \citep{waves3} & 2.5-2.67   &  $-$  &   $-$  &  $13.58_{-}^{+0.27}$  &  $14.23_{-0.10}^{+0.04}$  &   $14.65_{-0.03}^{+0.02}$  \\[0.15cm]
\hline
\end{tabular}}
 \end{minipage}
\end{table*}
\begin{table*}[!htp]
\centering
\caption{The predicted radii of a few high mass compact stars  in $f(Q,T)$ gravity (from Fig. \ref{f7}).}\label{tab3}
\begin{minipage}{0.8\textheight}
 \scalebox{0.9}{\begin{tabular}{| *{7}{c|} }
\hline
 &    & \multicolumn{5}{c|}{Predicted $R$ (km) ~~$f(Q,T)$}  \\[0.15cm]
\cline{3-7}
{Objects} & {$\frac{M}{M_\odot}$} & \multicolumn{5}{c|}{ $\psi_2$}  \\[0.15cm]
\cline{3-7}
&  & $0$ & $0.25$ & $0.50$ & $0.75$ & $1.0$  \\[0.15cm] \hline
PSR J074+6620 \citep{r1-PSRJ1810+1744-1} & 2.08$\pm$0.07  & $-$  &  $-$  &   $-$  &  $-$  &  $-$     \\[0.15cm]
\hline
PSR J1810+1744 \citep{r1-PSRJ1810+1744-1} & 2.13$\pm$0.04  & $-$  &  $-$  &   $-$  &  $-$  &  $-$     \\[0.15cm]
\hline
PSR J1959+2048 \citep{star1} & 2.18$\pm$0.09  & $-$  &  $-$  &   $-$  &  $-$  &  $-$    \\[0.15cm]
\hline
PSR J2215+5135 \citep{star1} & $2.28^{+0.10}_{-0.09}$ &  $14.68_{-}^{+0.01}$  &  $-$  &   $-$  &  $-$  &  $-$    \\[0.15cm]
\hline
GW190814 \citep{waves3} & 2.5-2.67 &    $14.65_{-0.03}^{+0.02}$  &  $14.72_{-0.04}^{+0.02}$  &  $14.79_{-0.04}^{+0.03}$  &   $14.88_{-0.05}^{+0.02}$ &  $14.97_{-0.05}^{-}$    \\[0.15cm]
\hline
\end{tabular}}
 \end{minipage}
\end{table*}
\begin{table*}[!htp]
\centering
\caption{The predicted radii of a few high mass compact stars  in $f(Q,T)$ gravity (from Fig. \ref{f7}).}\label{tab4}
\begin{minipage}{0.8\textheight}
 \scalebox{0.9}{\begin{tabular}{| *{7}{c|} }
\hline
 &    & \multicolumn{5}{c|}{Predicted $R$ (km) ~~$f(Q,T)$}  \\[0.15cm]
\cline{3-7}
{Objects} & {$\frac{M}{M_\odot}$} & \multicolumn{5}{c|}{$\beta$} \\[0.15cm]
\cline{3-7}
&   & $0$ & $0.25$ & $0.50$ & $0.75$ & $1.0$ \\[0.15cm] \hline
PSR J074+6620 \citep{r1-PSRJ1810+1744-1} & 2.08$\pm$0.07    &   $13.74_{-0.06}^{+0.04}$  &  $13.72_{-0.08}^{+0.04}$  &   $13.70_{-0.06}^{+0.04}$  &  $13.68_{-0.06}^{+0.05}$  &  $13.67_{-0.06}^{+0.04}$   \\[0.15cm]
\hline
PSR J1810+1744 \citep{r1-PSRJ1810+1744-1} & 2.13$\pm$0.04    &   $13.70_{-0.11}^{+0.03}$  &  $13.68_{-0.04}^{+0.03}$  &   $13.66_{-0.04}^{+0.03}$  &  $13.64_{-0.04}^{+0.04}$  &  $13.63_{-0.04}^{+0.03}$   \\[0.15cm]
\hline
PSR J1959+2048 \citep{star1} & 2.18$\pm$0.09     &  $13.65_{-0.12}^{+0.08}$  &   $13.63_{-0.08}^{+0.08}$  &  $13.61_{-0.13}^{+0.08}$  &  $13.59_{-0.13}^{+0.08}$ &   $13.58_{-0.13}^{+0.08}$ \\[0.15cm]
\hline
PSR J2215+5135 \citep{star1} & $2.28^{+0.10}_{-0.09}$ &   $13.51_{-0.37}^{+0.13}$  &  $13.48_{-0.42}^{+0.14}$  &   $13.46_{-}^{+0.14}$  &  $13.44_{-}^{+0.14}$  &  $13.42_{-}^{+0.15}$  \\[0.15cm]
\hline
GW190814 \citep{waves3} & 2.5-2.67   &  $-$  &   $-$  &  $-$  &  $-$  &   $-$  \\[0.15cm]
\hline
\end{tabular}}
 \end{minipage}
\end{table*}

In an another research work \cite{Becerra:2024wku}, mass-radius relations have been studied  for various EOSs and for varying anisotropic parameter. This work explore the stability of massive stellar systems against the radial perturbations are more preferable for the positive values of anisotropic parameter than the negative values. In the present investigation, we have considered positive and finite values of $\beta$ lying within $\{0,1\}$. The research work \cite{Becerra:2024wku} have shown that the massive  secondary  companion of  GW190814 can not be explained by vanishing anisotropic parameter or the isotropic model but the anisotropic systems  fulfilling different imposed conditions for a particular EOS such as BHF and GM1Y6 can explain the secondary  companion of  GW190814  with anisotropic parameter larger than 0.56 for the EOS GM1Y6. On the other hand, the bottom panel of Fig. \ref{f7} depicts that the secondary  companion of  GW190814 can not be explained in the present investigation for any value of $\beta$ within $\{0,1\}$. 

Based on the present analysis of the $M-R$ curves shown in the panels of Fig. \ref{f6}  and \ref{f7}, we have tabulated the predicted radii for the observed stars in the Tables \ref{tab1}, \ref{tab2}, \ref{tab3} and \ref{tab4}. In particular, the radii predicted in case of $f(Q)$ gravity as well as in $f(Q,T)$ gravity for varying parameter $\psi_1$ are shown Tables \ref{tab1} and \ref{tab2} respectively in order to compare the results in both the gravity theories. The results obtained for both $f(Q)$ gravity and  $f(Q,T)$ gravity apparently differ from each other for $\psi_1>1.00$. It is shown that the range of radius is \{10.5 km, 14.5 km\} for the EOSs considered in  a  study \cite{Saha:2024swd}. The Tables \ref{tab1} and \ref{tab2} suggest that the predicted radii for the observed stars of different mass fall within the aforementioned range \{10.5 km, 14.5 km\} for $\psi_1\leq 1.05$. The Tables \ref{tab3} and \ref{tab4} present the predicted radii for the massive stars in $f(Q,T)$ gravity for varying $\psi_2$ and $\beta$ respectively.  Again the radius of PSR J074+6620 is predicted to fall within \{13.09 km, 14.66 km\} which is in agreement with the predicted radii range \{11.79 km, 15.01 km\} found in \cite{Dittmann:2024mbo}.

\section{Concluding remarks} \label{VI}
In the present study, we have investigated the effect of model parameters $\{\psi_1, \psi_2\}$ and anisotropic parameter ($\beta$) on the effective density ($\rho^{eff}$) and effective pressures, the mass-radius relation and stability criterion of highly massive and dense compact stars in the context of $f(Q, T)$-gravity taking observational constrains  associated to the observed pulsars into account. The assumption of a physically valid metric potential and anisotropy function enable us to obtain solution in a closed-form to the modified field equations in $f(Q, T)$-gravity. The observed pulsars such as PSR J074+6620 \cite{r1-PSRJ074+6620}, PSR J1810+1744 \cite{r1-PSRJ1810+1744-1}, PSR J1959+2048 \cite{star1}, PSR J2215+5135 \cite{star1}, and merger event GW190814 \cite{waves3}  are explained in the present gravitational model which is found to be physically viable in accordance with the observational data. 

By matching the solution to the interior metric potentials with an exterior metric of Schwarzschild-de Sitter spacetime, we have obtained the unknown constants of the present model in terms of anisotropy parameter. With respect to varying values of model parameters $\{\psi_1, \psi_2, \beta\}$ we have analyzed the physical properties, stability condition and mass-radius relation of the present stellar configuration. This analysis have established the physical acceptance of the current gravitational stellar model in $f(Q,T)$ gravity. At first, the effective density of the present stellar system have been evaluated on the basis of its physical behavior from center to the surface for values of varying parameters $\{\psi_1, \psi_2, \beta\}$ as shown in the three panels respectively in Fig \ref{f1}. It has non-zero and finite values at the center and decreasing nature towards the surface for different values of the model parameters. It is  inferred that non-metricity has a vital role to play in the high density stars in the background of $f(Q,T)$ gravity similar to that in $f(Q)$ gravity \cite{Maurya:2024wtj,Maurya:2024dhk}. In particular, the central density as well as the effective density at each point in the star rise with a significant amount if one increases the value of $\psi_1$.  Now, increasing values of $\psi_2$ lead to decreasing values of effective density at every point inside the star. Additionally, the rate of change of the effective density with respect to $\psi_1$ at a particular radial coordinate is larger than that of $\psi_2$ at the same point.  So, left panel and middle panel of  Fig. \ref{f1} show a significant effect that impact of $\psi_1$ on effective density runs counter to the impact of parameter $\psi_2$ on the changes in the effective density \cite{Maurya:2025psm}. On the other hand, right panel of  Fig. \ref{f1} suggests that the increasing values of $\beta$ have an insubstantial effect on the positive change in the effective density everywhere inside the star.

When we have inspected both the effective pressures we see that the tangential pressure and radial pressure have identical variations throughout the star as this can be seen in Fig. \ref{f2}. Interestingly, both the effective pressures have  exactly the same value at the center of the star leading to the zero central anisotropy and the vanishing of radial pressure at the surface of the star is fulfilled for different values of model parameters. The tangential pressure are slightly larger than the radial pressure near the surface as it can be seen from the panels of Fig. \ref{f2}. Both the pressures increase more in magnitude around the central region than that in magnitude near the surface for increasing values of parameters $\psi_1$ and $\psi_2$. However, the pressures show decreasing trend at the central region for increasing $\beta$. 

As $\beta$ approaches to zero the pressures in radial and tangential direction  tend to become equal in magnitude and identical in nature which in agreement with the fact that the anisotropy becomes zero for the vanishing $\beta$ as indicated in the expression of $\Delta^{\text{eff}}$ and illustrated graphically in the panels of Fig. \ref{f3}.  Further, the anisotropy shows increasing nature with regard to the radial distance for various values of the parameters while its value becomes maximum at the surface. The impact of parameter $\psi_1$ on the anisotropy is more prominent as compared to the effect of parameter $\psi_2$ on the anisotropy. Now, $\beta$  tunes the anisotropy within the stellar structure as shown in the right panel of Fig. \ref{f3}. 

The nature of change in the sound speeds with respect to radial distance and varying parameters has been investigated graphically in the panels of Fig. \ref{f4} which confirms that the present stellar system in the framework of $f(Q,T)$ gravity follows the causality condition, i.e., $0\leq v^2_r \leq 1$ and  $0\leq v^2_t \leq 1$  everywhere inside the star under the variation $f(Q,T)$ parameters and anisotropy parameter. Then we checked the adiabatic condition for stability analysis by showing the graphical nature of $\Gamma$ with regard to radial distance in the three panels of Fig. \ref{f5} corresponding to the variations in model parameters. The adiabatic index has increasing trend within the anisotropic stellar system as shown in each panel with   no direct influence of the non metricity  as interpreted by the left panel of Fig. \ref{f5}.  In the present anisotropic system, the central $\Gamma$ is greater than three and it increases with high slope confirming that the present stellar system is stable under the adiabatic condition. The increasing values of $\psi_2$ enhance the value of $\Gamma$ everywhere inside the star as indicated from the middle panel in Fig. \ref{f5}. On the other hand, increasing values of anisotropy parameter decrease the values of $\Gamma$ for each radial distance as depicted in the right panel in Fig. \ref{f5}.     

We have investigated the modified Tolman-Oppenheimer-Volkoff (TOV) equation for $f(Q,T)$ gravity and have analyzed the hydrostatic equilibrium balance of the present stellar configuration under  $F_g$, $F_h$, $F_a$ and $F_{(Q,T)}$  the forces representing gravitational, hydrostatic, anisotropic and non-minimal coupled type interactions respectively. We have examined these forces with respect to radial distance in the panels of Fig. \ref{f5tov} for varying parameters. We get an important result that hydrostatic and anisotropic forces are  repulsive  whereas gravitational and coupling forces are  attractive. Subsequently,  the contrary nature of these forces stabilizes the stellar  by neutralizing the net combined effect of these forces.  

Additionally, we have investigated the Harrison-Zel'dovich-Novikov criterion  by showing the variation of total mass with respect to central density  for the varying parameters in the three panels of Fig. \ref{f5mro}. We obtained the result from  each panel that total mass has a monotonically increasing nature with regard to the central density. The impact of the parameters on the $M-\rho_c$ curves is seen to be nominal in magnitude. So,  the present stellar configuration attain stable equilibrium under small radial perturbations following the Harrison-Zel'dovich-Novikov criterion,i.e., $dM/d\rho_c >0$.   

We have studied the impact of anisotropy and the model parameters on the properties of highly dense stars and mass-radius relation in the context of $f(Q,T)$ gravity by considering some observed massive pulsars. The mass-radius curves are represented in Fig.~\ref{f6} for various values of $\psi_1$ with constant value of anisotropy parameter where the top panel  and bottom panel of Fig.~\ref{f6} implies the $f(Q)$ gravity and $f(Q,T)$ gravity respectively showing a the impact of $\psi_1$ on $MR$ curves is similar for both cases for $\psi_1\leq1.0$. One of the important result we obtain is that $MR$ curves for $\psi_1>1.0$ can not predict the radius for less massive stars because of the increase in nuclear density beyond the maximum limit in case of $f(Q,T)$ gravity   signifying the less massive stars tends to have its maximum masses attaining cores with higher densities relative to the cores of comparatively highly massive stars \cite{Saha:2024swd}. In both the panels of Fig. \ref{f6}, the $M-R$ curves the peak or $M_{max}$ and the corresponding radius increase due to increase in $\psi_1$.  The mass-radius relation with regard to varying $\psi_2$ for fixed values of $\psi_1$ and $\beta$ has been represented in the top panel of Fig. \ref{f7}. The $M_{max}$ from the each $MR$ curve remains almost same for increasing $\psi_2$ whereas the radius corresponding to the $M_{max}$ increases as shown in the Figure \ref{f7}. The values of $M_{max}$ obtained from the peak of $MR$ curves shown in Fig. \ref{f7} are seen to be increasing for decreased values of $\beta$. This implies that the increased amount of anisotropy decreases $M_{max}$ of the present stellar configuration with nominal effect of $\beta$ on the radius corresponding to $M_{max}$ in the framework of $f(Q,T)$ gravity. By comparing the particular curve of $MR$ in  $f(Q)$ gravity and the $MR$ curve with $\beta=0$ in $f(Q,T)$ gravity, we see that the stellar structure with higher compactness and higher density for zero anisotropy is more preferable in the background of $f(Q)$ gravity relative to that in  $f(Q,T)$ gravity.

In the present investigation, we have considered positive and finite values of $\beta$ lying within $\{0,1\}$  \cite{Becerra:2024wku}. In this research work authors have shown that the massive  secondary  companion of  GW190814 can  be explained the anisotropic systems  fulfilling different imposed conditions for a particular EOS such as BHF and GM1Y6  in particular with anisotropic parameter larger than 0.56 for the EOS GM1Y6. On the other hand, the bottom panel of Figure \ref{f7} depicts that the secondary  companion of  GW190814 can not be explained in the present investigation for any value of $\beta$ within $\{0,1\}$. Based on the present analysis of the $MR$ curves shown in the panels of Figure \ref{f6}  and \ref{f7}, we have tabulated the predicted radii for the observed stars in the Tables \ref{tab1}, \ref{tab2}, \ref{tab3} and \ref{tab4}. It is shown that the range of radius is \{10.5 km, 14.5 km\} for the EOSs considered in  a  study \cite{Saha:2024swd}. The Tables \ref{tab1} and \ref{tab2} suggests that the predicted radii for the observed stars of different mass fall within the aforementioned range \{10.5 km, 14.5 km\} for $\psi_1\leq 1.05$. Again the radius of PSR J074+6620 is predicted to fall within \{13.09 km, 14.66 km\} which is in agreement with the predicted radii range \{11.79 km, 15.01 km\} found in \cite{Dittmann:2024mbo}.

In summary, while the present study successfully describe the properties of  high dense massive pulsars with respect to $f(Q,T)$ parameters and anisotropy consistent with astrophysical data and GW observations, there remains considerable scope for further research to examine the applicability and limitations of the alternative gravity theories in context of the third generation GW observational data \cite{Punturo:2010zza, Evans:2021gyd}.

\section*{Acknowledgments}
The author SKM acknowledges that the Ministry of Higher Education, Research, and Innovation (MoHERI) supported this research work through the project BFP/GRG/CBS/24/035 and he also thankful to UoN administration for continuous support and encouragement for the research works.

\section*{Data Availability Statement}
This manuscript has no associated data or the data will not be deposited, as such there is no observational data related to this article where the necessary calculations and graphic discussion are already available in the manuscript.

\section*{Conflict of Interest}
The authors declare that they have no conflict of interest or personal relationships that could have appeared to influence the work reported in this paper.

\newpage

\section*{Appendix}
\begin{eqnarray}
    && \mathcal{C}_3=\Bigg[48 (\mathcal{B}+1) \pi  (\mathcal{B}+\sigma +1) \bigg(24 \left(\sigma ^3+4 \sigma ^2-32 \sigma +128\right) \left(2 \mathcal{B}^2+(\sigma +14) \mathcal{B}+3 \sigma -4\right) \nonumber\\&& \hspace{0.5cm}\times\tanh ^{-1}\left(\frac{\mathcal{B} (\sigma -4)+5 \sigma +12}{\sqrt{7-\mathcal{B} (\mathcal{B}+10)} \sqrt{16-(\sigma -8) \sigma }}\right) \sqrt{-\mathcal{B}^2-10 \mathcal{B}+7} (\mathcal{B}+\sigma +1)^3+\left(\mathcal{B}^2+10 \mathcal{B}-7\right)\nonumber\\&& \hspace{0.5cm}\times \Big(21 \sigma ^6-150 \sigma ^5+1218 \sigma ^4-748 \sigma ^3-11248 \sigma ^2-15680 \sigma +4 \mathcal{B}^4 \left(\sigma ^3+10 \sigma ^2+80 \sigma -288\right)\nonumber\\&& \hspace{0.5cm}+2 \mathcal{B}^3 \left(7 \sigma ^4+56 \sigma ^3+564 \sigma ^2-448 \sigma -5952\right)+2 \mathcal{B}^2 \big(9 \sigma ^5+43 \sigma ^4+660 \sigma ^3+1780 \sigma ^2-12800 \sigma \nonumber\\&& \hspace{0.5cm}-13632\big)+\mathcal{B} \left(9 \sigma ^6+12 \sigma ^5+490 \sigma ^4+4816 \sigma ^3-16456 \sigma ^2-44160 \sigma -23424\right)-6912\Big) \nonumber\\&& \hspace{0.5cm} \times \sqrt{-\sigma ^2+8 \sigma +16}\bigg)+\Bigg(24 \sqrt{-\mathcal{B}^2-10 \mathcal{B}+7} (\mathcal{B}+\sigma +1)^3 \left(\sigma ^3+4 \sigma ^2-32 \sigma +128\right) \big(9 \mathcal{B}^4\nonumber\\&& \hspace{0.5cm}+(66-5 \sigma ) \mathcal{B}^3+\left(-4 \sigma ^2-79 \sigma +120\right) \mathcal{B}^2+\left(-32 \sigma ^2+137 \sigma +78\right) \mathcal{B}+36 \sigma ^2+51 \sigma +15\big) \nonumber\\&& \hspace{0.5cm} \times \tanh ^{-1}\left(\frac{\mathcal{B} (\sigma -4)+5 \sigma +12}{\sqrt{7-\mathcal{B} (\mathcal{B}+10)} \sqrt{16-(\sigma -8) \sigma }}\right)-\left(\mathcal{B}^2+10 \mathcal{B}-7\right) \sqrt{-\sigma ^2+8 \sigma +16} \Big(-18 (\sigma ^3\nonumber\\&& \hspace{0.5cm}+10 \sigma ^2+80 \sigma -288) \mathcal{B}^6-4 \left(11 \sigma ^4+80 \sigma ^3+904 \sigma ^2+480 \sigma -13824\right) \mathcal{B}^5-2 \big(8 \sigma ^5-82 \sigma ^4+277 \sigma ^3\nonumber\\&& \hspace{0.5cm}+9738 \sigma ^2-12720 \sigma -103968\big) \mathcal{B}^4+3 (15 \sigma ^6+132 \sigma ^5+1544 \sigma ^4+496 \sigma ^3-55776 \sigma ^2+76288 \sigma \nonumber\\&& \hspace{0.5cm}+133632) \mathcal{B}^3+(15 \sigma ^7+783 \sigma ^6-1708 \sigma ^5+19352 \sigma ^4-133766 \sigma ^3+100068 \sigma ^2+689952 \sigma \nonumber\\&& \hspace{0.5cm}+430272) \mathcal{B}^2+(174 \sigma ^7-1209 \sigma ^6+6212 \sigma ^5-59820 \sigma ^4+15536 \sigma ^3+506048 \sigma ^2+673152 \sigma \nonumber\\&& \hspace{0.5cm}+244224) \mathcal{B}+3 (-75 \sigma ^7+183 \sigma ^6-2940 \sigma ^5-2900 \sigma ^4+28502 \sigma ^3+71244 \sigma ^2+62048 \sigma \nonumber\\&& \hspace{0.5cm}+19008)\Big)\Bigg) \psi _2\Bigg]\Bigg/\Big[6 \sqrt{-\mathcal{B}^2-10 \mathcal{B}+7} (\mathcal{B}+\sigma +1)^3 \left(-\sigma ^2+8 \sigma +16\right)^{7/2} \Big(48 (\mathcal{B}+1) \pi  (2 \mathcal{B}^3\nonumber\\&& \hspace{0.5cm}+(3 \sigma +16) \mathcal{B}^2+\left(\sigma ^2+18 \sigma +10\right) \mathcal{B}+3 \sigma ^2-\sigma -4)+\big(9 \mathcal{B}^4+(66-5 \sigma ) \mathcal{B}^3+(120-4 \sigma ^2-79 \sigma)\nonumber\\&& \hspace{0.5cm}\times \mathcal{B}^2+\left(-32 \sigma ^2+137 \sigma +78\right) \mathcal{B}+36 \sigma ^2+51 \sigma +15\big) \psi _2\Big)\Big]^{-1}. \nonumber
\end{eqnarray}


\end{document}